\documentclass[peerreview,journal,draftcls,onecolumn,12pt]{IEEEtran}
\usepackage{hyperref}
\usepackage{graphicx}
\usepackage{epstopdf}
\usepackage{amssymb}
\usepackage{amsmath}
\usepackage{multirow}
\usepackage{multicol}
\usepackage{subfigure}
\usepackage[ruled,vlined]{algorithm2e}
\usepackage{bbm}
\usepackage{array}
\usepackage{empheq}
\usepackage{dblfloatfix}
\usepackage{placeins}
\usepackage{booktabs}
\usepackage{color}
\usepackage{amsthm}

\newtheorem{theorem}{\textbf{Theorem}}
\newtheorem{lemma}{\textbf{Lemma}}

\newtheorem{corollary}{\textbf{Corollary}}
\newtheorem{definition}{\textbf{Definition}}
\newtheorem{remark}{\textbf{Remark}}

\newcolumntype{?}{!{\vrule width 1pt}}

\begin{document}
\title{Latency-Optimal Uplink Scheduling Policy in Training-based Large-Scale Antenna Systems}
\author{Kyung~Jun~Choi,~\IEEEmembership{Student~Member,~IEEE},~and~Kwang~Soon~Kim$^\dag$,~\IEEEmembership{Senior~Member,~IEEE}
\thanks{}
\thanks{
This work was supported in part by ICT R\&D program of MSIP/IITP
[B0126-16-1012, Multiple Access Technique with Ultra-Low
Latency and High Efficiency for Tactile Internet Services in IoT
Environments], and in part by Basic Science Research Program through the National Research Foundation of Korea (NRF) funded by the Ministry of Education, Science and Technology (NRF-2014R1A2A2A01007254).

The authors are with the Department of Electrical and Electronic Engineering, Yonsei University, 50 Yonsei-ro, Seodaemun-gu, Seoul, 03722, Korea.}
\thanks{$^\dag$: Corresponding author (ks.kim@yonsei.ac.kr)}
}
\markboth{Choi and Kim: Latency-Optimal Uplink Scheduling Policy in Training-based Large-Scale Antenna Systems}{in preparation} 
\maketitle
\vspace{-1cm}
\begin{abstract}
In this paper, an uplink scheduling policy problem to minimize the network latency, defined as the air-time to serve all of users with a quality-of-service (QoS), under an energy constraint is considered in a training-based large-scale antenna systems (LSAS) employing a simple linear receiver. An optimal algorithm providing the exact latency-optimal uplink scheduling policy is proposed with a polynomial-time complexity. Via numerical simulations, it is shown that the proposed scheduling policy can provide several times lower network latency over the conventional ones in realistic environments. In addition, the proposed scheduling policy and its network latency are analyzed asymptotically to provide better insights on the system behavior. Four operating regimes are classified according to the average received signal quality, $\rho$, and the number of BS antennas, $M$. It turns out that orthogonal pilots are optimal only in the regime $\rho\gg1$ and $ M\ll \log^2\rho$. In other regimes ($\rho\ll 1$ or $ M\gg \log^2\rho$), it turns out that non-orthogonal pilots become optimal. More rigorously, the use of non-orthogonal pilots can reduce the network latency by a factor of $\Theta(M)$ when $\rho\ll 1$ or by a factor of $\Theta(\sqrt{M}/\log M)$ when $\rho\gg 1$ and $M\gg \log\rho$, which would be a critical guideline for designing 5G future cellular systems.
\end{abstract}
\section{Introduction}

\IEEEPARstart{D}{ue} to continuous introduction of mobile devices and services, future cellular systems are facing a significantly increased number of mobile devices requesting large data volume. To accommodate such a large growth of mobile devices, there are active researches on the 5th generation (5G) cellular system. New targets for the 5G cellular system are to support latency-sensitive applications such as Tactile Internet \cite{TactileInternet} and low energy consumption for machine-type communication (MTC) \cite{TalebMTC} or the Internet of things (IoT) \cite{AtzoriIoT}. Unfortunately, a cellular system cannot achieve the two targets simultaneously, but a non-trivial tradeoff can exist.
Although this tradeoff is very important to 5G cellular system designers, related researches are rare. This is because it is often hard to deal with the latency and the energy consumption analytically so that intensive simulation-based network plannings are widely spread \cite{IkunoSLS}, \cite{PiroSimulating}. However, this approach becomes impractical when the network components, such as the number of users and BS antennas are scaled up. More viable approach is to analyze the network. This paper mainly concentrates on the analysis about the tradeoff between the latency and the energy consumption in a promising 5G cellular system. 

In 5G cellular systems, there has been great interest to a large-scale antenna system (LSAS), a.k.a. massive multiple-input multiple-output (MIMO), in which very large number of antennas are equipped at a base station (BS) to serve many users simultaneously \cite{MarzettaNoncooperative}. Its inherent merits come from massive spatial dimensions, which include i) generating sharp beams for intended users to improve spectral efficiency by suppressing unintended interference \cite{MarzettaNoncooperative}, \cite{RusekScaling}, ii) reducing transmit energy while guaranteeing quality of service (QoS) \cite{NgoEnergy}, and iii) allowing a complexity-efficient transceiver algorithm \cite{MullerLinear1}. In order to achieve such advantages, an appropriate channel state information (CSI) acquisition process is essential. To acquire CSI, a widely-accepted approach is the training-based transmission in which a frame is divided into two phases: one is the training phase, in which users transmit known training signals and the BS estimates the CSI, and the other is the data transmission phase, in which the users transmit information-bearing signals and the BS extracts the information by utilizing the estimated CSI. Even if the training-based transmission is not optimal in information-theoretic point of view, it gives an efficient way to acquire the CSI as well as to provide the optimal degrees of freedom in the high signal-to-noise ratio (SNR) regime \cite{ZhengCommunication}.

In order to analyze the latency in the training-based LSAS, it is necessary to optimize the user scheduling policy as well as the resource allocation under reasonable and practical constraints. If this optimization is not performed, it often gives an inappropriate cellular system design. The optimization of the training-based transmission is firstly investigated by Hassibi and Hochwald \cite{Hassibi}. They consider the MIMO point-to-point channel with a capacity-approaching transmitter/receiver pair and successfully derive the optimal system parameters as a function of SNR and other parameters. Later, this results are extended to the MIMO broadcast channel \cite{MaryBC}, multiple access channel \cite{MurugesanMAC}, relay channel \cite{SunRelay}, and interference channel \cite{AyachIC}. However, these works optimize the energy and time dedicated to the training phase only under a given user set so that it cannot be directly applied to the latency-energy tradeoff in the LSAS. In order to evaluate the latency of the LSAS, it is necessary to further optimize those variables under the optimal scheduling policy. 

The scheduling policies to minimize the latency (or delay) under a minimum rate constraint or to maximize spectral efficiency under a maximum latency constraint have been widely investigated in literature under various system models. In \cite{IdoOptimal}, the system average delay is optimized by using combined energy/rate control under average symbol-energy constraints. In \cite{LauDelay}, delay-optimal energy and subcarrier allocation is proposed for orthogonal frequency division multiple access (OFDMA). In \cite{RajanDelay}, the energy minimizing scheduler, by adapting energy and rate based on the queue and channel state is proposed. However, most of them assume perfect CSI at transmitter and receiver so that it often overestimates the network-wise performance. Also, their scheduling policies are too complicated to be analyzed for providing an intuitive insight on the network-wise performance. Thus, a practically optimal scheduling policy for the training-based LSAS is needed and an intuitive analysis is desired to provide an insight on the latency-energy tradeoff in the LSAS.

Decreasing the latency in the LSAS is closely related to increasing the spectral efficiency, because higher spectral efficiency results in a smaller transmission completion time if the number of users and their rate constraints are given. In addition, the spectral efficiency of a multiple-access channel with $M$ BS antennas and $K$ scheduled users is asymptotically expressed as $\min\{M,K\}\log(\mathsf{SNR})+O(1)$ as $\mathsf{SNR}\to \infty$, which implies that the spectral efficiency can be enhanced by scheduling users as many as possible in the LSAS. However, most literature assumes that \emph{orthogonal} pilots are allocated to users so that the maximum number of scheduled users is limited by the number of available pilots in practice. Actually, there is no reason that orthogonal pilots are optimal for the latency-energy tradeoff so that it is natural to consider \emph{non-orthogonal} pilots in general. There are a few results related to the case using non-orthogonal pilots. In \cite{WangDesign}, optimal non-orthogonal pilots for minimizing channel estimation error are derived and it turns out that finding the optimal non-orthogonal pilots is equivalent to solving the Grassmannian subspace packing problem. In \cite{ShenDownlink}, an iterative algorithm is proposed to find optimal non-orthogonal pilots for maximizing the number of users with a minimum rate constraint in a downlink LSAS. However, they still do not address the effect of the non-orthogonal pilots on the latency and it would be very interesting to find whether the use of non-orthogonal pilots can reduce the latency and when and how much reduction can be obtained over the case of using orthogonal pilots. 

In this paper, we are interested in an uplink training-based LSAS serving many users with an average energy constraint, in which each user has a limited average energy for transmitting a frame. In addition, we assume a block Rayleigh fading model and a practical receiver such as the maximum ratio combining (MRC) or the zero-forcing (ZF) receiver and focus on the resource allocation and multiple access strategy (to be specific, pilot allocation, user grouping and scheduling, and energy allocation). The main target of this paper is to address the following question: \textbf{\emph{how much time is needed for guaranteeing the minimum throughput to all users in the uplink training-based LSAS?}} It is, in general, hard to address this question analytically so that we look into the two asymptotic regimes, high and low energy regimes and successfully derive the effect on the network latency according to the energy consumption. 

The main contributions of this paper are summarized as follows:

\begin{itemize}
\item We optimize the uplink scheduling policy for minimizing the latency with guaranteeing the minimum rate constraint. The optimizing variables are the scheduling groups in which users are simultaneously scheduled, the scheduling portion indicating how often each scheduling group actually transmits, and the energy allocation indicating how much portion of energy is dedicated to the training phase. This problem is transformed into an equivalent problem of maximizing the spectral efficiency with the rate constraint. The optimal scheduling policy is obtained by solving the binary integer programming (BIP) and it is proved that the optimal solution of the original BIP can be obtained by a linear programming relaxation with a polynomial-time complexity. 

\item We investigate the asymptotic performance of the proposed optimal uplink scheduling policy for a large number of users. We derive a simple close-form expression for the asymptotic network latency and find the optimal parameters for the proposed optimal uplink scheduling policy. Then, we identify four operating regimes of the training-based LSAS according to the growth or decay rate of the average received signal quality, $\rho$, and the number of BS antennas, $M$.  It turns out that orthogonal pilot sequences are optimal only when the average received signal quality is sufficiently good and the number of BS antennas is not-so-large. In other regimes, it turns out that non-orthogonal pilot sequences become optimal. 
In fact, the use of non-orthogonal pilots can reduce the network latency by a factor of $\Theta(M)$ when the received signal quality is quite poor ($\rho\ll 1$) or by a factor of $\Theta(\sqrt{M}/\log M)$ when the received signal quality is sufficiently good ($\rho\gg 1$) and the number of BS antennas is sufficiently large ($M\gg \log_2\rho$).

\end{itemize}

The remainder of this paper is organized as follows. In Section II, a detailed model description is presented including channel, energy, and signal models for a training-based LSAS. In Section III, the uplink scheduling policy problem is formulated and its optimal solution is provided. Numerical experiments to verify the superiority of the proposed uplink scheduling policy are shown in Section VI. 
In Section V, an asymptotic analysis provides the closed-form network latency and optimal parameters for the proposed uplink scheduling policy. Finally, conclusion is drawn in Section VI.

Matrices and vectors are respectively denoted by boldface uppercase and lowercase characters. Also, $(\cdot)^T$, ${( \cdot )^H}$, and $\left| \cdot \right|$ stand for the transpose, conjugate transpose, and cardinality of a set, respectively, and $\log$ and $\log_2$ are the natural logarithm and the logarithm with base 2, respectively. Also, $\lceil x \rfloor$ denotes the function rounding $x$ towards the nearest integer, $(x)^+ = \max\{x,0\}$, and $\mathbbm{1}\{\cdot\}$ denotes the indicator function. 
$\mathcal{CN}(\boldsymbol{\mu},\mathbf{R})$ denotes the distribution of a circularly symmetric complex Gaussian random vector with mean vector $\boldsymbol{\mu}$ and covariance matrix $\mathbf{R}$ and $\mathbb{E}[ \cdot ]$ and $\mathrm{Var}(\cdot)$ denotes the statistical expectation and the statistical variance, respectively. Finally, standard order notations in \cite{KnuthBig} are used. For better readability, frequently used symbols are summarized in Table I.

\begin{figure}[t]
\centering
\includegraphics[width=0.5\columnwidth]{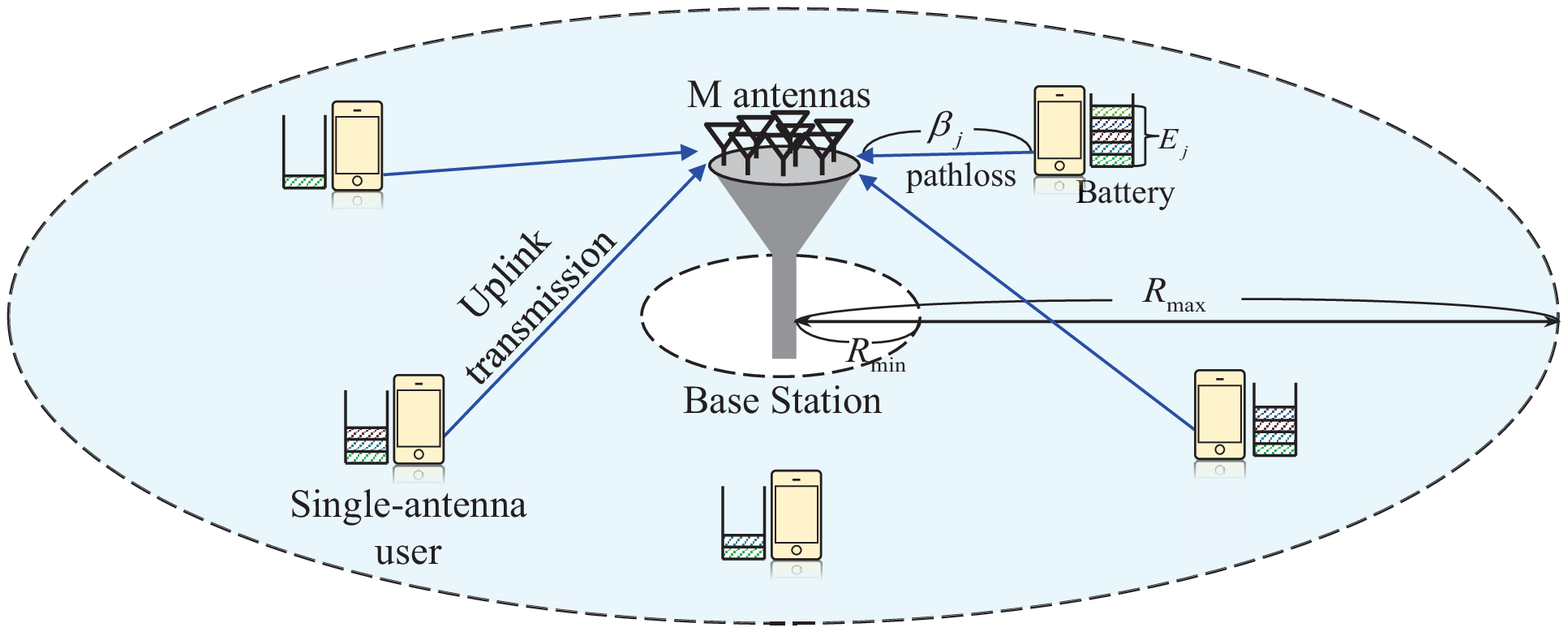}
\caption{System model. }
\end{figure}
\begin{figure}[t]
\centering
\includegraphics[width=0.5\columnwidth]{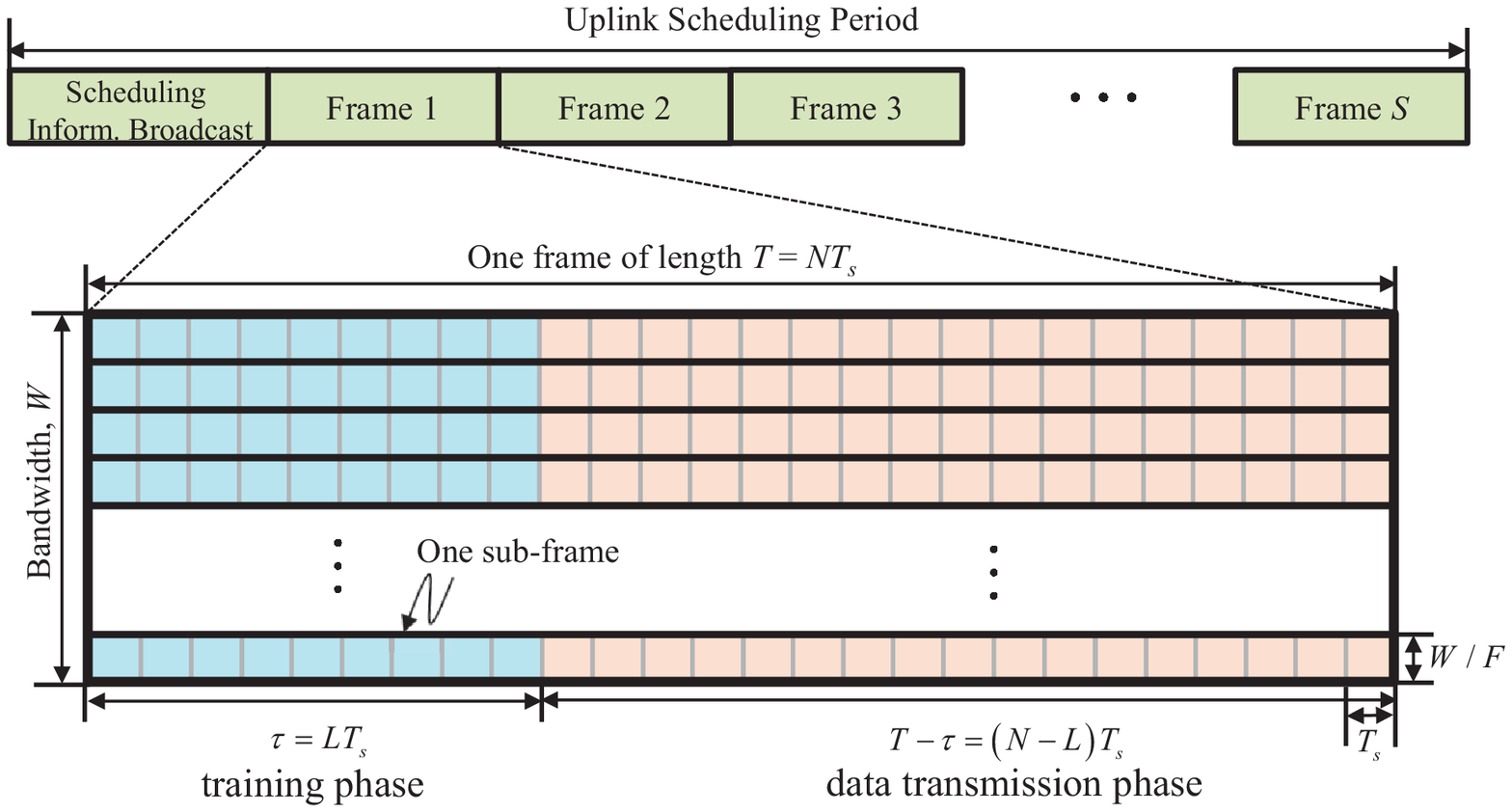}
\caption{Frame model.}
\end{figure}

{\linespread{1}
\renewcommand{\arraystretch}{1.1}
\begin{table}[t]
\centering
\caption{Symbol Notation Summary}
\begin{tabular}{?c|m{6cm}?}
\specialrule{.1em}{.05em}{.05em} 
Symbols& Descriptions\\
\hline\hline
$M$ & \# of BS antennas \\
\hline
$U,~ \mathcal{U}$ & \# of total users and their index set, $\mathcal{U}=\{1,\dots,U\}$\\
\hline
$S,~S^\star$ & \# of frames in a uplink scheduling period and its minimum with guaranteeing $\mathcal{T}_{\rm{th}}$ bits to all of users. \\
\hline
$~\mathcal{D}^\star,~\overline{\mathcal{D}}^\star$ & network latency, optimized network latency and optimized asymptotic network latency normalized by $U$ with guaranteeing $\mathcal{T}_{\rm{th}}$ bits to all of users. \\
\hline
$F$ & \# of sub-frames in a frame \\
\hline
$N$ & \# of symbols in a sub-frame\\
\hline
$L,~N-L$ & \# of symbols in the training phase or the data transmission phase\\
\hline
$W,~T_s$ & bandwidth and the symbol duration\\
\hline
$\eta$ & bandwidth inefficiency with $\eta = WT_s/F \ge 1$\\
\hline
$E_j$ & the average energy constraint of user $j$.\\
\hline
$\widetilde{R}_k[f;t], {R}_k[f;t] $ & the achievable rate of users $j$ in sub-frame $f$ of frame $t$ and its approximation (bits/Hz)\\
\hline
$\mathcal{T}_{\rm{th}}$& throughput threshold (bits)\\
\hline
$\mathcal{O}_q,~D_q$ & the scheduling group and its scheduling portion\\
\hline
$p_j^{\rm{tr}},~p_j^{\rm{dt}}$ & the energy dedicated to the training symbols or data symbols (Joule/symbol)\\
\specialrule{.1em}{.05em}{.05em} 
\end{tabular}
\end{table}
}

\section{System Model}
We consider an uplink LSAS consisting of a BS with $M$ antennas, and $U$ single-antenna users as illustrated in Fig. 1. It is assumed that the users are randomly distributed on the cell coverage region and they want the quality of service (QoS) on their own data traffic (rate, latency, and reliability) so that the BS serves these users persistently and try to guarantee their QoS. 

A two-phase frame structure with training and data transmission phases, illustrated as in Fig. 2, is used. For every uplink scheduling period, the BS broadcasts the scheduling information and then the users transmit $S$ frames in uplink direction step by step. The frame of time length $T$ seconds and bandwidth $W$ Hz is divided into $F$ equal-bandwidth sub-frames by partitioning frequency domain by using the orthogonal division multiplex access (OFDMA), single-carrier frequency domain multiple access (SC-FDMA) or any good one of the newly considered waveforms \cite{BanelliModulation}. The sub-frame consists of $N$ symbols of time period $T_s$ seconds. In the training phase of time period $\tau=LT_s$ seconds, the scheduled users send $L$ training symbols, and the BS estimates the uplink channels. Then, in the data transmission phase of the remaining time period $T-\tau=(N-L)T_s$ seconds, all of the scheduled users transmit $N-L$ data symbols to the BS simultaneously in a space division multiple access (SDMA) manner.\footnote{Here, $F$ sub-frames and $N$ symbols in the sub-frame can be arbitrarily configured both in time and frequency domains. For example, in the Long-Term Evolution (LTE), one sub-frame includes $168$ symbols (14 symbols in time domain and 12 sub-carriers in frequency domain) and one frame includes $10$ sub-frames in time domain. } So, the transmit signal vector of user $j$, who is allocated to the $f$th sub-frame in frame $t$, is written as 
$${{\bf{x}}_j}[f;t] = {\left[ {{{\left( {{\bf{x}}_j^{{\rm{tr}}}[f;t]} \right)}^H},{{\left( {{\bf{x}}_j^{{\rm{dt}}}[f;t]} \right)}^H}} \right]^H},$$ 
where $\mathbf{x}_j^{\rm{tr}}[f;t]$ is the $L\times 1$ training symbol vector for the training phase and $\mathbf{x}_j^{\rm{dt}}[f;t]$ is the $(N-L)\times 1$ data symbol vector for the data transmission phase.

In every sub-frame, at most $M$ users are scheduled and the set of users scheduled in sub-frame $f$ of frame $t$ is denoted as $\mathcal{S}[f;t]$. Assume that $|\mathcal{S}[f;t]|=K$.
Within one block of $N$ symbols, the $M\times N$ received signal matrix, denoted as $\mathbf{Y}[f;t]=\left[\mathbf{y}_1[f;t],\mathbf{y}_2[f;t],\dots,\mathbf{y}_N[f;t]\right]$, is given as
\begin{equation}\label{eq1}
\begin{split}
\mathbf{Y}[f;t]
&= \mathbf{G}_{\mathcal{S}[f;t]}[f;t]\mathbf{X}^H[f;t] + \mathbf{V}[f;t],
\end{split}
\end{equation}
where $\mathbf{G}_{\mathcal{S}[f;t]}[f;t] = \left[\mathbf{g}_{j}[f;t]\right]_{j\in\mathcal{S}[f;t]}$ is the $M\times K$ channel matrix, $\mathbf{X}[f;t]=[\mathbf{x}_1[f;t],\dots,\mathbf{x}_K[f;t] ]$ is the $N\times K$ transmitted signal matrix and $\mathbf{V}[f;t]=\left[\mathbf{v}_{1}[f;t],\mathbf{v}_{2}[f;t],\dots,\mathbf{v}_{N}[f;t]\right]$ is the $M\times N$ noise matrix, whose elements are independent and identically distributed (i.i.d.) random variables with $\mathcal{CN}(0,1)$.\footnote{Note that since the noise power is normalized to unity, the transmit power is in fact the relative power with respect to noise power.}
The $M\times 1$ flat-fading channel vector between the BS and user $j$ at the sub-frame $f$ of frame $t$, $\mathbf{g}_{j}[f;t]$, can be written as
\begin{equation}\label{eq2} 
{{\mathbf{g}}_{j}}[f;t] = \sqrt {{\beta _{j}}} {{\mathbf{h}}_{j}}[f;t],
\end{equation}
where ${{\mathbf{h}}_{j}}[f;t] \in {\mathbb{C}^{M \times 1}}$ is the short-term CSI whose elements are i.i.d. random variables with $\mathcal{CN}(0,1)$ and ${\beta _{j}}( \ge 0)$ is the long-term CSI depending on the path-loss and shadowing. The long-term CSI between the BS and user $j$ is modeled as $\beta_{j} = d_{j}^{-\alpha}$, where $\alpha (> 2)$ is the wireless channel path-loss exponent and $d_j$ is the distance between the BS and user $j$. We assume that Rayleigh block fading model, where the short-term CSI of each user remains constant within a given frame but is independent across different frames, while the long-term CSI does not vary during a much longer interval. Further, it is assumed that the long-term CSI of all users is perfectly known at the BS.

Since each of users has a different limited battery capacity, recharge process, or radio frequency (RF) transmitter, they are assumed to be limited to spend energy for transmitting each sub-frame. Let ${E}_j$ be the average allowed energy level (in Joule) of user $j$ per sub-frame of length $T$. The energy is consumed during both the training phase of length $\tau$ and the data transmission phase of length $T-\tau$. So, the consumed energy transmitting each sub-frame needs to meet $$\mathbb{E}\|\mathbf{x}_j[f;t]\|^2 = \mathbb{E}\|\mathbf{x}_j^{\rm{tr}}[f;t]\|^2+\mathbb{E}\|\mathbf{x}_j^{\rm{dt}}[f;t]\|^2\le E_j,~\forall j.$$ Letting $p_j^{\rm{tr}} = \mathbb{E}\|\mathbf{x}_j^{\rm{tr}}[f;t]\|^2/L$ and $p_j^{\rm{dt}}=\mathbb{E}\|\mathbf{x}_j^{\rm{dr}}[f;t]\|^2/(N-L)$ be the transmit energy of each training symbol and data symbol, respectively, the constraint is represented as
\begin{equation}\label{eq3}
\frac{L}{N} p_j^{\rm{tr}} +\left(1-\frac{L}{N}\right)p_j^{\rm{dt}} \le \frac{{E}_j}{N},~\forall j.
\end{equation}
In the sequel, we drop the sub-frame index $f$ and the frame index $t$ if there is no ambiguity.

\subsection{Training Phase}
To estimate the CSI of $K$ scheduled users, the BS allocates $K$ pilot sequences with length of $L$. 
Let $\boldsymbol{\Psi}_{L,K} = [\boldsymbol\psi _1,\boldsymbol\psi_2,\dots,\boldsymbol\psi _K]$ be the $L\times K$ pilot matrix with normalized columns, i.e., $\|\boldsymbol{\psi}_j\|^2 = 1$ for all $j$. 
The training symbol vector of scheduled user $j$ during the training phase is $\mathbf{x}_j^{\rm{tr}} = \sqrt{Lp_j^{\rm{tr}}}\boldsymbol\psi_j$. For equalizing the difference of all users' channel estimation quality (maximizing the worst), the \emph{channel-inversely power-controlled pilots} are assumed similarly as in \cite{ShenDownlink}, in which the average received signal energy of the users in $\mathcal{S}[f;t]$ is set to the common target received energy, $\overline{{p}^{\rm{tr}}}$. So, the transmit energy at the training phase is set by 
\begin{equation}\label{eq4}
p_j^{\rm{tr}} = {\beta^{-1}_{j}}{\overline{p^{\rm{tr}}}}.
\end{equation}
Then, the $M\times L$ received signal matrix in frame $t$ at the BS during the training phase, denoted as ${{\mathbf{Y}}}^{\rm{tr}} = \left[\mathbf{y}_{1},\mathbf{y}_{2},\dots,\mathbf{y}_{L}\right]$, can be written as
\begin{align}\label{eq5}
{{\mathbf{Y}}}^{\rm{tr}}&= \sqrt {L\overline{p^{{\rm{tr}}}}} {{\mathbf{H}}}_{\mathcal{S}[f;t]}\boldsymbol\Psi _{L,K}^H + {{\mathbf{V}}}^{\rm{tr}},
\end{align}
where $\mathbf{H}_{\mathcal{S}[f;t]} = \left[\mathbf{h}_{j}\right]_{j\in\mathcal{S}[f;t]}$, and ${{\mathbf{V}}}^{\rm{tr}} = \left[\mathbf{v}_{1},\mathbf{v}_{2},\dots,\mathbf{v}_{L}\right]$ is the noise vector during the training phase.
Using the minimum mean-square error (MMSE) channel estimator \cite{LiPilot}, the estimated short-term CSI of user $j$, denoted as $\widehat {\bf{h}}_j$, can be written as
\begin{equation}\label{eq6}
\begin{split}
\widehat {\bf{h}}_j & = \sqrt {L\overline {{p^{{\rm{tr}}}}} } {\bf{Y}}^{\rm{tr}}{\left( {{\bf{I}}_{L} + L\overline {{p^{{\rm{tr}}}}} {\bf{\Psi }}_{L,K}{{\bf{\Psi }}_{L,K}^H}} \right)^{ - 1}}{\boldsymbol{\psi }}_j.
\end{split}
\end{equation}
Denote the channel estimation error by $\widetilde{\mathbf{h}}_{k}=\widehat{\mathbf{h}}_{k}-\mathbf{h}_{k}$. The following lemma informs the property of the MMSE channel estimation. 
\begin{lemma}
With the channel-inversely power-controlled pilots, 
$\widehat{\mathbf{h}}_{k}\sim\mathcal{CN}(\mathbf{0},(1-\sigma_{\rm{tr}}^2)\mathbf{I}_M)$ and $\widetilde{\mathbf{h}}_{k}\sim\mathcal{CN}(\mathbf{0},\sigma_{\rm{tr}}^2\mathbf{I}_M)$ are mutually independent and the channel estimation error variance $\sigma_{\rm{tr}}^2$ is given by 
\begin{align}\label{eq7}
\sigma _{{\rm{tr}}}^2 &={{{{\left(1 - \frac{L}{K}\right)}^ + }}} + \frac{1}{K}\sum\nolimits_{i = 1}^{\min \{ L,K\} } {\frac{1}{{1 + L\overline {{p^{{\rm{tr}}}}} {\lambda _i}}}} ,
\end{align}
where $\lambda_i$ are the eigenvalues of $\mathbf{\Psi}_{L,K}^H\mathbf{\Psi}_{L,K}$.
\begin{IEEEproof}
See the proof of Theorem 1 in \cite{WangDesign}.
\end{IEEEproof}
\end{lemma}


\subsection{Data Transmission Phase}
During the data transmission phase, the $l$th received signal vector at sub-frame $f$ of frame $t$, $l=L+1,\dots,N$, is given by 
\begin{equation}\label{eq8}
\mathbf{y}_l[f;t] = \sum\limits_{j\in\mathcal{S}[f;t]} \sqrt{p_j^{\rm{dt}}\beta_j}\mathbf{h}_j s_{jl}[f;t] + \mathbf{v}_l,
\end{equation}
where $s_{jl}[f;t]\sim\mathcal{CN}(0,1)$ is the $l$th information-bearing data symbol of user $j$ at sub-frame $f$ of frame $t$.

By treating the estimated CSI as if it were the true CSI, the BS selects a linear receiver $\mathbf{F}$ such as the zero-forcing (ZF) receiver or maximum ratio combining (MRC), given by 
\begin{align}\label{eq9}
\mathbf{F} = \left\{\begin{array}{*{20}{l}}
{\widehat {\bf{G}}_{\mathcal{S}[f;t]}{{\left( {{{\widehat {\bf{G}}}^H_{\mathcal{S}[f;t]}}\widehat {\bf{G}}_{\mathcal{S}[f;t]}} \right)}^{ - 1}}},&\text{ if ZF},\\
\widehat {\bf{G}}_{\mathcal{S}[f;t]},&\text{ if MRC},
\end{array}\right.
\end{align}
where $\widehat {\bf{G}}_{\mathcal{S}[f;t]}$ is the estimated version of ${\bf{G}}_{\mathcal{S}[f;t]}$.
Such a linear receiver becomes nearly optimal in the LSAS, i.e, $M\gg K $\footnote{$f(n)\ll g(n)$ means that $\lim_{n\to\infty}f(n)/g(n) = 0 $, i.e., $f(n)=o(g(n))$.} \cite{RusekScaling}. Then, the $l$th signal of user $k\in\mathcal{S}[f;t]$ after using the linear receiver can be written as
\begin{equation}\label{eq10}
\begin{split}
{r_{kl}}[f;t] =& \sqrt {p_k^{{\rm{dt}}}{\beta _k}} {\bf{f}}_k^H{\widehat {\bf{h}}_k}{s_{kl}}[f;t] + \sum\limits_{i \in {\cal S}[f;t]\backslash \{ k\} } {\sqrt {p_i^{{\rm{dt}}}{\beta _i}} {\bf{f}}_k^H{{\widehat {\bf{h}}}_i}{s_{il}}}[f;t] \\
&~~~~~~ - \sum\limits_{j \in {\cal S}[f;t]} {\sqrt {p_j^{{\rm{dt}}}{\beta _j}} {\bf{f}}_k^H{{\widetilde {\bf{h}}}_j}{s_{jl}}}[f;t] + {\bf{f}}_k^H{\bf{v}}_l,
\end{split}
\end{equation}
where $\mathbf{f}_k$ denotes the $k$th column vector of $\mathbf{F}$. In (\ref{eq10}), the first term is the desired signal, the second term is the inter-user interference, the third term is the interference caused from the imperfect channel estimation, and the last term is the noise. Note that the second term disappears when the ZF receiver is applied.
Treating the interference as Gaussian random variables, the achievable rate of user $k\in\mathcal{S}[f;t]$ (bits/symbol) during sub-frame $f$ of frame $t$ is given by 
\begin{equation}\label{eq11}
{\widetilde{R}_{k}}[f;t] = \log_2\left( {1 + {\widetilde{\gamma}_{k}[f;t]}} \right),
\end{equation}
where $\widetilde{\gamma}_{kl}[f;t]$ is the signal-to-interference-plus-noise-ratio (SINR), given by
\begin{equation}\label{eq12}
\begin{split}
&{\widetilde{\gamma}_{k}}[f;t] = \frac{p_k^{{\rm{dt}}}{\beta _k}{{{\left| {{\bf{f}}_k^H{{\widehat {\bf{h}}}_k}} \right|}^2}}}{{{{\left\| {{{\bf{f}}_k}} \right\|}^2}\left( {1 + \sigma _{{\rm{tr}}}^2\sum\limits_{j \in {\cal S}[f;t]} {p_j^{{\rm{dt}}}{\beta _j}} } \right) + \sum\limits_{j \in {\cal S}[f;t]\backslash \{ k\} } {p_j^{{\rm{dt}}}{\beta _j}{{\left| {{\bf{f}}_k^H{{\widehat {\bf{h}}}_j}} \right|}^2}} }}.
\end{split}
\end{equation}
Since the use of the exact distribution of (\ref{eq11}) is analytically intractable\footnote{Even if the perfect CSI is provided, $\widetilde{\gamma_k}$ follows a Chi-square distribution so that $\mathbb{E}[\log(1+\widetilde{\gamma_k})]$ involves a hyper-geometric function which makes the exact analysis intractable \cite{GoreZF}.}, the following Lemma is used instead.

\begin{lemma} With the channel-inversely power-controlled non-orthogonal pilots, the achievable rate of user $k$ of the MRC or ZF receiver is approximated by 
\begin{equation}\label{eq13}
{R_k}[f;t] \approx \left\{ {\begin{array}{*{20}{l}}
{\log_2\left( {1 + \gamma _k[f;t]} \right),}&{{\text{if }}k \in {\cal S}[f;t],}\\
{0,}&{{\text{if }}k \notin {\cal S}[f;t],}
\end{array}} \right.
\end{equation}
 where $\gamma_{k}[f;t]$ is given as
{
\begin{equation}\label{eq14}
\begin{split}
&{\gamma_{k}}[f;t]=  \left\{ {\begin{array}{*{20}{l}}
{\frac{{\left( {1 - \sigma _{{\rm{tr}}}^2} \right)\left( {M - K} \right)p_k^{{\rm{dt}}}{\beta _k}}}{{1 + \sigma _{{\rm{tr}}}^2\sum\limits_{j \in {\cal S}[f;t]} {p_j^{{\rm{dt}}}{\beta _j}} }}},&{{\text{if ZF}}},\\
{\frac{{\left( {1 - \sigma _{{\rm{tr}}}^2} \right)\left( {M - 1} \right)p_k^{{\rm{dt}}}{\beta _k}}}{{1 + {\sum\limits _{j \in \mathcal{S}[f;t]}}p_j^{{\rm{dt}}}{\beta _j} - \left( {1 - \sigma _{{\rm{tr}}}^2} \right)p_k^{{\rm{dt}}}{\beta _k}}}},&{{\text{if MRC}}}.
\end{array}} \right.
\end{split}
\end{equation}
}
\end{lemma}
\begin{IEEEproof}
The proof is similar as in \cite{NgoEnergy}, so omitted for brevity.
\end{IEEEproof}

Note that (\ref{eq14}) is valid only when $M\ge K-1$ for the ZF receiver ($M\ge 2$ for the MRC receiver). In fact, when such an approximation is used as the utility function of a scheduler, it cannot select $K=M$ users, which under-utilizes the resource, especially when $M$ is small. However, this paper deals with the LSAS so that this problem becomes less significant. Additionally, to overcome this problem for a small $M$, $M-K$ can be replaced by $M-K+1$, which results in a small approximation error. (Similar approximation is shown in Theorem 5 in \cite{ZhangPower}.) Furthermore, this approximated version of the achievable rate becomes more accurate as the number of BS antennas increases.


\begin{remark} [Rate achievability]
Since the low latency communications are of interest, codewords cannot be spread sufficiently in time domain. Also, the transmitter does not acquire its instantaneous CSI, the transmission strategy cannot be adapted according to the channel realization. In this paper, we assume that user $j$ transmits information-bearing data symbols with a fixed rate of $R_k$ regardless of the channel realization. The BS can successfully receive this data symbol if and only if $\widetilde{R}_k \ge R_k$. From the strong law of large number, we can show that $\widetilde{R}_k \ge R_k$ almost surely as $M\to\infty$ so that the rate $R_k$ is achievable in the LSAS with high probability. This is called \emph{channel hardening effect} of the LSAS, which implies that the instantaneous rate is approaching a deterministic quantity, i.e., its mean, as $M\to \infty$. See \cite{HochwaldChannelHardening} for more rigorous discussions. 
\end{remark}

To maximize the average achievable rate, we need to design pilot sequences.
\begin{lemma}
Optimal pilot sequences $\Psi_{L,K}$ for maximizing $R_k$ satisfies 
\begin{equation}\label{eq15}
\mathbf{\Psi}_{L,K}\mathbf{\Psi}_{L,K}^H = {\max\left\{1,\frac{K}{L}\right\}}\mathbf{I}_{L}.
\end{equation}
\end{lemma}
\begin{IEEEproof}
Since the case $L\ge K$ is trivial so we omit. Let $L<K$. The channel estimation error variance (\ref{eq7}) is simplified into 
\begin{equation*}
 \sigma_{\rm{tr}}^2 = 1 - \frac{1}{K}\sum\limits_{i = 1}^L {\left( {\frac{{L\overline {{p^{{\rm{tr}}}}} {\lambda _i}}}{{1 + L\overline {{p^{{\rm{tr}}}}} {\lambda _i}}}} \right)}. 
\end{equation*}
Obviously, $\gamma_k^{\rm{lb}}$ is a decreasing function with $\sigma_{\rm{tr}}^2$ within $0 \le \sigma_{\rm{tr}}^2\le 1$.
So, the maximization of $\gamma^{\rm{lb}}_k$ is equivalent to the maximization $\sum\nolimits_{i = 1}^L {\left( {\frac{{L\overline {{p^{{\rm{tr}}}}} {\lambda _i}}}{{1 + L\overline {{p^{{\rm{tr}}}}} {\lambda _i}}}} \right)}$ under $\sum\nolimits_{i=1}^L\lambda_i = K$, which is obtained when the eigenvalues are the same, i.e., $\lambda_i = K/L$.
\end{IEEEproof}

Note that the above condition is identical to the condition minimizing the channel estimation error \cite{WangDesign}. In fact, this condition is known as the Welch bound equality (WBE) \cite{KangChu}. So, from Lemma 3, the channel estimation error variance is minimized at
\begin{equation}\label{eq16}
\sigma_{\rm{tr}}^2 = \frac{{1 + {{\left( {K - L} \right)}^ + }\overline {{p^{{\rm{tr}}}}} }}{{1 + \max \{ L,K\} \overline {{p^{{\rm{tr}}}}} }},
\end{equation}
and the SINR is maximized at 
\begin{equation}\label{eq17}
\begin{split}
&{\gamma_k}[f;t] = \left\{ {\begin{array}{*{20}{l}}
{\frac{{L\left( {M - K} \right)p_k^{{\rm{dt}}}{\beta _k}\overline {{p^{{\rm{tr}}}}} }}{{\left( {1 + {{\left( {K - L} \right)}^ + }\overline {{p^{{\rm{tr}}}}} } \right)\left( {1 + \sum\limits_{j \in \mathcal{S}[f;t]} {p_j^{{\rm{dt}}}{\beta _j}} } \right) + L\overline {{p^{{\rm{tr}}}}} }},}&{{\text{if ZF}},}\\
{\frac{{L\left( {M - 1} \right)p_k^{{\rm{dt}}}{\beta _k}\overline {{p^{{\rm{tr}}}}} }}{{\left( {1 + \max \left\{ {L,K} \right\}\overline {{p^{{\rm{tr}}}}} } \right)\left( {1 + \sum\limits_{j \in \mathcal{S}[f;t]} {p_j^{{\rm{dt}}}{\beta _j}} } \right)-L\overline{p^{\rm{tr}}}p_k^{\rm{dt}}\beta_k}},}&{{\text{if MRC}}.}
\end{array}} \right.
\end{split}
\end{equation}
If the pilot sequences are under-utilized (orthogonal pilots are used), i.e., $K\le L$, $\sigma_{\rm{tr}}^2$ is simplified into $\sigma_{\rm{tr}}^2 = 1/(1+L\overline{p^{\rm{tr}}})$, which means that the channel estimation error depends only on the energy dedicated to the training phase $L\overline{p^{\rm{tr}}}$. However, if the pilot sequences are over-utilized (non-orthogonal pilots are used), i.e., $K>L$, additional interference $(K-L)\overline{p^{\rm{tr}}}$ occurs.

\begin{remark} [Optimal non-orthogonal pilots] 
Obviously, if $K\le L$, the optimal pilot sequences can be obtained from arbitrarily chosen $K$ columns of an $L\times L$ unitary matrix. In the case of $K>L$, one option is to over-sample an $L\times L$ unitary matrix. Although there are infinitely many sequences holding the WBE, one simple example is the discrete Fourier transform (DFT) based sequences, obtained as
\begin{equation*}
{\mathbf{\Psi} _{L,K}} = \frac{1}{{\sqrt L }}\left[ {\begin{array}{*{20}{l}}
1&{{e^{ - \mathsf{j}\frac{{2\pi {f_1}}}{{ K }}}}}& \dots &{{e^{ - \mathsf{j}\frac{{2\pi {f_1}(K - 1)}}{{ K }}}}}\\
1&{{e^{ - \mathsf{j}\frac{{2\pi {f_2}}}{{ K }}}}}& \dots &{{e^{ - \mathsf{j}\frac{{2\pi {f_2}(K - 1)}}{{ K }}}}}\\
 \vdots & \vdots & \ddots & \vdots \\
1&{{e^{ - \mathsf{j}\frac{{2\pi {f_L}}}{{ K }}}}}& \dots &{{e^{ - \mathsf{j}\frac{{2\pi {f_L}(K - 1)}}{{ K }}}}}
\end{array}} \right],
\end{equation*}
where $\mathsf{j} = \sqrt{-1}$ and $0<f_1<f_2<\dots<f_L<K$ are arbitrarily chosen integers.
Note that the DFT-based sequences are widely used in the design of the unitary-space time modulation \cite{HochwaldUSTM} or the feedback codebook \cite{XiaWelchBound}.
\end{remark}

The accuracy of the approximated rate in (\ref{eq13}) and (\ref{eq17}) is presented in Fig. 3, by setting $N=100$, $L=10$, $K=5$, $E\in\{0,10,20\}$dB, $p_j^{\rm{dt}} = \frac{E}{2(N-L)}$, $p_j^{\rm{tr}} = \frac{E}{2L}$, and $\beta_j = 1$, $\forall j$. In Fig. 3, we plot the exact rate in (\ref{eq12}) and the approximated rate in {(\ref{eq13})} according to the number of BS antennas, $M$. The dashed line represents the average of the exact rate and each error bar represents its standard deviation. We can easily notice that the approximated rate is quite well fitted even at a small number of antennas in a wide range of average allowed energy level $E$.

\begin{figure}
\centering
\includegraphics[width=0.5\columnwidth]{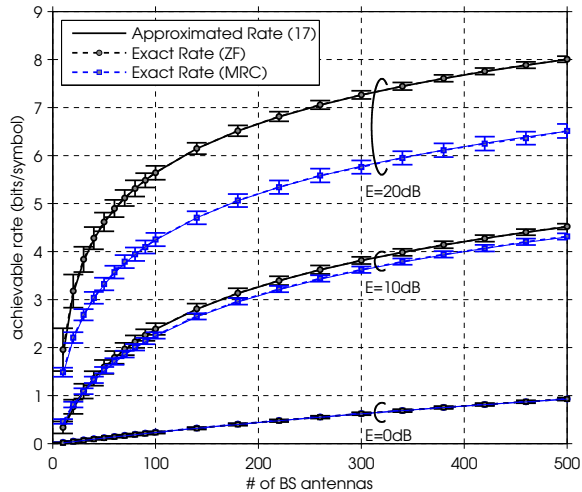}
\caption{Accuracy of the approximated rate in (\ref{eq13}) and (\ref{eq17}). The line marked by a circle represents the results using the ZF receiver and the line marked by a square represents the results using the MRC receiver.}
\end{figure}

\section{Optimal Static Uplink Scheduling Policy}
\subsection{Static Uplink Scheduling Policy}
According to the level of accessible information, uplink scheduling policies can be classified into two types. A \emph{dynamic} user scheduling, which is based on the \emph{instantaneous} CSI, can provide a substantial rate gain primarily because it allows the BS to select a subset of users whose channels are nearly orthogonal and the achievable uplink performance increases with the number of users. However, obtaining the instantaneous CSI at a BS will incur a large amount of uplink training resource cost. In fact, due to the nature of the limited channel coherence time, it is hard for every user to participate in the scheduling pool without incurring non-negligible overhead. On the contrary, a \emph{static} uplink scheduling exploits the long-term information only, such as the CSI statistics and/or the average allowed energy levels, which can be easily acquired via infrequent feedback with negligible overhead. 

A static scheduling policy for $S$ frames is defined as $(\mathcal{O}, \mathcal{D}, \mathcal{P},L)$, where 
$\mathcal{O}=\{\mathcal{O}_1, \mathcal{O}_2,\dots, \mathcal{O}_Q\}$ denotes the set of scheduling groups, 
$\mathcal{D}=\{D_1,D_2,\dots,D_Q\}$ denotes the set of scheduling portions, 
$\mathcal{P} =\{(p_1^{\rm{tr}},p_1^{\rm{dt}}),(p_2^{\rm{tr}},p_2^{\rm{dt}}),\dots,(p_U^{\rm{tr}},p_U^{\rm{dt}})\}$ denotes the set of energy allocations for all users, and $L$ determines the sub-frame configuration.
Due to the nature of the static uplink scheduling, the output of the static scheduling policy should satisfy the following two constraints: 1) $\mathcal{O}_p \bigcap \mathcal{O}_q = \emptyset$ if $\forall p\ne q$, and 2) $\bigcup_{q=1}^Q \mathcal{O}_q = \mathcal{U}$, where $\mathcal{U}=\{1,2,\dots,U\}$. Also, the scheduling portion 
is defined as 
\begin{equation}\label{eq18}
D_q = \frac{1}{FS}\sum\nolimits_{t=1}^S\sum\nolimits_{f=1}^F \mathbbm{1}{\{ \mathcal{S}[f;t] = \mathcal{O}_q\}},\end{equation} so that it must satisfy $\sum_{q=1}^Q D_q = 1$. 

Note that $D_q$ is the portion of the sub-frames allocated to scheduling group $q$, $\mathcal{O}_q$, during one scheduling period consisting of $FS$ sub-frames. In fact, since $D_qFS$ is not an integer, $\lfloor{D_qFS\rfloor}$ sub-frames may be allocated. We assume that $SF$ is sufficiently large so that the error $\lfloor{D_qFS\rfloor} - D_qFS$ is negligibly small. 
{
\linespread{1}
\begin{table}[t]
\centering
\caption{Values According To Receiver Types}
\scalebox{0.8}{
\begin{tabular}{?c?c|c?}
\specialrule{.1em}{.05em}{.05em} 
{Values}&{ZF Receiver}&{MRC Receiver}\\
\hline
\hline
{${a}_{K,L} $}&{$
 L\left({M - K} \right){E_K}{\beta _K}$}&{$
 L\left( {M - 1} \right){E_K}{\beta _K}$}\\
\hline
{${b}_{K,L}$}&{$
 {L^2(M-K)}$}&{$
L^2\left( {M - 1} \right) $}\\
\hline
{${c}_{K,L} $}&{$
 K{E_K}{\beta _K} + N - L $}&{$
 K {{E_K}{\beta _K}}+N - L $}\\
\hline
{${d}_{K,L} $}&{$
\left( {N - L - K} \right)L + {\left( {K - L} \right)^ + }{c_{K,L}}$}
 &{$ \max \left\{ {L,K} \right\}c_{K,L}- KL-LE_K\beta_K $}\\
\hline
{${e}_{K,L}$}&{$ {KL(K-L)^+}
$}&{$L({K\max \left\{ {L,K} \right\} }-L) $}\\
\specialrule{.1em}{.05em}{.05em} 
\end{tabular}}
\end{table}
}

\subsection{Latency Minimization Problem Formulation}
In general, a user tries to send some of information to the BS via a wireless medium within various constraints. However in cellular systems, the limited wireless medium is shared so that severe delay often occurs for waiting transmission turns and also for transmitting the target data volumes. Thus, to guarantee a low latency is one of the main hurdles to be addressed in future cellular systems. 
\begin{definition}

The network latency (delay) is defined as $\mathcal{D}^\star = {T}S^\star$ (sec), where $S^\star$ is the minimum number of frames required to deliver the target throughput of $\mathcal{T}_{\rm{th}}$ bits for each user.\footnote{Although the queue dynamics is another delay source, the latency caused from the shared wireless channel is considered only by assuming zero queuing delay.}
\end{definition}

{Our definition is different to \cite{SharifDelay}, \cite{XiaNonOrthogonal}, in which the delay is defined as the scheduling delay (the waiting time for transmission turns) only and the delay for transmitting the target throughput volume is ignored, and also to \cite{LiuCompletion}, \cite{CalabuigMAC}, in which the delay is defined as the transmission completion time for the target throughput only and ignores the effect of the scheduling delay. Our definition includes both scheduling delay and transmission completion time.}

To minimize the latency under a given throughput constraint $\mathcal{T}_{\rm{th}}$, it is sufficient to minimize the number of frames, $S$. Thus, the optimization problem can be constructed as follows:
\begin{subequations}\label{eq19}
\begin{empheq}[box=\fbox]{align}
(\text{P}) &\underset{\mathcal{O},\mathcal{D}, \mathcal{P},L}{\text{min}}~~S,~~ \text{subject to}& \\
&(N-L)\sum\nolimits_{t=1}^{S} \sum\nolimits_{f=1}^F R_j[f;t]\ge \mathcal{T}_{\rm{th}}, \forall j,&\\
&{\frac{L}{N}} {p_j^{\rm{tr}}} \!\!+\!\!\left(\!1\!-{\frac{L}{N}}\right)p_j^{\rm{dt}} \le \frac{E_j}{N},~{p_j^{\rm{tr}}}, p_j^{\rm{dt}}\ge 0, \forall j,&\\
&\mathcal{O}_p \cap \mathcal{O}_q = \emptyset,~\bigcup\nolimits_{p=1}^{Q}\mathcal{O}_p= \mathcal{U},&\\
&\sum\nolimits_{q=1}^Q D_q = 1,~D_q \ge 0, \forall q,&\\
&L\in\{1,2,\dots,N-1\}.&
\end{empheq}
\end{subequations}
The constraint (\ref{eq19}b) is to guarantee the required throughput $\mathcal{T}_{\rm{th}}$ for each of users, (\ref{eq19}c) is to meet the average energy constraint, and (\ref{eq19}d)-(\ref{eq19}e) are to meet the static scheduler condition. 
The variables in the optimization problem (P) are as follows: 
\begin{itemize}
\item[1)] Which of users are simultaneously scheduled? ($\mathcal{O}=\{\mathcal{O}_1,\mathcal{O}_2,\dots,\mathcal{O}_Q\}$)
\item[2)] How many sub-frames are allocated to each group? ($\mathcal{D}=\{D_1,D_2,\dots,D_Q\}$)
\item[3)] How the energy is allocated to the training and data transmission phases? ($\mathcal{P} = \{(p_1^{\rm{tr}}, p_1^{\rm{dt}}),\dots,\newline(p_U^{\rm{tr}}, p_U^{\rm{dt}})\}$ for $\forall j$)
\item[4)] How much the symbols are dedicated to the training phase? ($L$)
\end{itemize}
The problem (P) is obviously non-convex and is very complicated, but can be transformed into an equivalent problem.

\subsection{Problem Transformation and Optimal Solutions}
Since we deal with a static uplink scheduling policy, the achievable rate of user $j$ is independent to the sub-frame index $f$ and frame index $t$ if user $j$ is scheduled. Thus, we use the notation $R_j$ instead of $R_j[f;t]$ if $j\in \mathcal{S}[f;t]$.
Let $j\in \mathcal{O}_q$ and define the spectral efficiency of user $j$ (in bps/Hz) as the average rate of user $j$ served by the BS normalized by time and bandwidth, given as
\begin{align}
{\mathsf{SE}}_{j} &=\frac{N-L}{WTS}\sum\limits_{t = 1}^{{S}} {\sum\limits_{f = 1}^F {{{R_j}[f;t]} } } = \frac{1-{\frac{L}{N}}}{\eta }D_q R_j,
\label{eq20}
 \end{align}
where $\eta = {WT_s}/{F}\ge 1$ denotes the bandwidth inefficiency (such as the cyclic prefix overhead), and the second equality comes from (\ref{eq18}). Since every user has the same throughput constraint $\mathcal{T}_{\rm{th}}$, it is required that $R_j \ge \Omega_q$ for $j \in \mathcal{O}_q$, where $\Omega_q$ is the common rate for scheduled group $q$. Inserting $R_j \ge \Omega_q$ into (\ref{eq20}), we have 
\begin{align}\label{eq21}
{\mathsf{SE}}_{j} &\ge {\mathsf{SE}}= \frac{1-{\frac{L}{N}}}{\eta}D_q\Omega_q.
\end{align}
In order to guarantee the same minimum rate for all users, we further set 
\begin{equation}\label{eq22}
D_q=\frac{{{ \Omega _q ^{ - 1}}}}{{\sum\nolimits_{i = 1}^Q {{\Omega _i ^{ - 1}}} }},~ \forall q.
\end{equation}
From (\ref{eq21}) and (\ref{eq22}), every user is guaranteed to exceed the common spectral efficiency 
\begin{equation}\label{eq23}
\mathsf{SE} = \frac{1-{\frac{L}{N}}}{\eta}\frac{1}{\sum\nolimits_{i=1}^Q \Omega_i^{-1}}.
\end{equation}

Our approach is first to maximize the spectral efficiency while providing the common spectral efficiency $\mathsf{SE}$ to every user in a cell at each possible value of $L$. To meet the target throughput $\mathcal{T}_{\rm{th}}$ for every user, the BS needs $S=\left\lceil \frac{\mathcal{T}_{\rm{th}}}{{W}T\mathsf{SE}}\right\rceil$ sub-frames so that the latency is given as 
\begin{equation}\label{eq24}
\mathcal{D}={T}\left\lceil \frac{\mathcal{T}_{\rm{th}}}{{W}T\mathsf{SE}}\right\rceil\approx  \frac{\mathcal{T}_{\rm{th}}}{W\mathsf{SE}},
\end{equation}
 where that last approximation is valid when $\frac{\mathcal{T}_{\rm{th}}}{{W}T\mathsf{SE}}$ is large. Then, an equivalent optimization problem can be formulated as follow:

\begin{subequations}\label{eq25}
\begin{empheq}[box=\fbox]{align}
\!(\text{P-eq})~& \underset{\mathcal{O},\mathcal{P}}{\text{maximize}} \!\!\!& \!&\! {\mathsf{SE}},\\
& \text{subject to} & & R_j\ge\Omega_q,~\forall j \in \mathcal{O}_q,~\forall q,\\
& & & \text{(\ref{eq19}c)}-\text{(\ref{eq19}d)}.
\end{empheq}
\end{subequations}
Note that a similar transformation is shown in \cite{TutuncuogluHarvesting}.
In the sequel, we devise the optimal uplink scheduling policy by solving (P-eq).

\subsubsection{Optimal Transmit Energy Allocation $\mathcal{P}^\star$ Under Given Users}
Assume that $L$ is fixed and users $\mathcal{O}_q=\{1,2,\dots, {K}\}$ are scheduled in the $q$th scheduling group and they are arranged in the descending order of $E_j\beta_j$, i.e., $E_1\beta_1 > E_2\beta_2 > \dots > E_{K}\beta_{K}$, without loss of generality. 
Note that each scheduled user should have non-zero transmit energy for both the training and data transmission phases. In order to find the optimal transmit energy allocation in (P-eq), the following sub-problem is considered.
\begin{subequations}\label{eq26}
\begin{align}
(\text{P-A})& \underset{\{p_j^{\rm{tr}}, p_j^{\rm{dt}}\}_{j\in\mathcal{O}_q}}{\text{maximize}}&& \Omega_q, &\\
& \text{subject to}& & \text{(\ref{eq19}c) and (\ref{eq25}b)}.&
\end{align}
\end{subequations}
To obtain the optimal solution, the following observations are helpful.
\begin{itemize}
\item Since $R_j$ is an increasing function of the transmit energy, the optimal transmit energy are obtained when $R_j = \Omega_q$ for $\forall j\in\mathcal{O}_q$ (no waste), i.e.,
\begin{equation}\label{eq27}
p_i^{\rm{dt},\star}\beta_i = p_j^{\rm{dt},\star}\beta_j,~\forall i,j \in \mathcal{O}_q. 
\end{equation}

\item Since $E_i\beta_i\ge E_j\beta_j$ for any $i<j$, if $p_j^{\rm{dt}} \ge p$ is feasible, then $p_i^{\rm{dt}} \ge p$ is also feasible. So, when $K$ users are scheduled, the optimal energy of the $K$th user should satisfy
\begin{equation}\label{eq28}
{p_K^{{\rm{dt}},\star} = \frac{{E_K{\beta _K} - L \overline{p^{{\rm{tr}}}}^\star}}{{\left( {N - L } \right){\beta _K}}}}.
\end{equation}
\end{itemize}
Under the assumption that $K$ users are scheduled and the above two observations, the objective function of (P-A) in (\ref{eq26}a) can be written as $\Omega_q = \overline{\Omega}_q(\overline{p^{\rm{tr}}};K,L)$ from (\ref{eq17}),
where 
\begin{equation}\label{eq29}
\overline{\Omega}_q(x;K,L) = {\log_2 }\left( {1 + \frac{{(a_{K,L} - b_{K,L}x )x }}{{c_{K,L}+ d_{K,L}x-e_{K,L}x^2}}} \right),
\end{equation}
with $a_{K,L}, b_{K,L}, c_{K,L}, d_{K,L}$ and $e_{K,L}$ are defined as in Table II.
Since the objective function of (P-A) is a single-variable function, it can be easily solved and the following theorem states the optimal transmit energy allocation.

\begin{theorem} 
For given $L$ and scheduling user set $\mathcal{O}_q$, the optimal transmit energy of user $j\in\mathcal{O}_q$ during the training and data transmission phases is given as 
\begin{align}\label{eq30}
\left(p_j^{{\rm{tr}}, \star },p_j^{{\rm{dt}}, \star }\right) ={\left(
\dfrac{{{u^\star }(\mathcal{O}_q,L)}}{{{\beta _j}}}, {\dfrac{{{E_{K}}{\beta _{K}} - L {u^ \star }(\mathcal{O}_q,L)}}{{\left( {N - L } \right){\beta _j}}}}
\right),}
\end{align}
where ${u^\star}(\mathcal{O}_q,L)$ is given in the bottom of the this page.
\begin{figure*}[b]
\hrulefill
\vspace*{4pt}
\begin{equation}\label{eq31}
\begin{split}
&{u^\star}(\mathcal{O}_q,L) =\\& \left\{ {\begin{array}{*{20}{l}}
{\dfrac{{{b_{K,L}}{c_{K,L}}}}{{{b_{K,L}}{d_{K,L}} - {a_{K,L}}{e_{K,L}}}}\left( {\sqrt {1 + \dfrac{{{a_{K,L}}}}{{{b_{K,L}}}}\dfrac{{{b_{K,L}}{d_{K,L}} - {a_{K,L}}{e_{K,L}}}}{{{b_{K,L}}{c_{K,L}}}}} - 1} \right),}&{{\text{if }}{b_{K,L}}{d_{K,L}} - {a_{K,L}}{e_{K,L}} \ne 0,}\\
{\dfrac{{{a_{K,L}}}}{{2{b_{K,L}}}},}&{{\text{if }}{b_{K,L}}{d_{K,L}} - {a_{K,L}}{e_{K,L}} = 0.}
\end{array}} \right.
\end{split}
\end{equation}
\end{figure*}
\end{theorem}
\begin{IEEEproof}
See Appendix A.
\end{IEEEproof}
For later use, we define 
\begin{equation}\label{eq32}
\Omega_q^\star(\mathcal{O}_q, L) = \overline{\Omega}(u^\star(\mathcal{O}_q,L);\left| \mathcal{O}_q\right|,L),
\end{equation}
as the optimal common rate for  given $\mathcal{O}_q$ and $L$. Note that by inserting (\ref{eq31}) into (\ref{eq29}) and using variables in Table II, it can be seen that $\Omega_q^\star(\mathcal{O}_q;L)$ is a non-decreasing function of $E_K\beta_K$ and is independent to $E_j\beta_j$, $\forall j\in\{1,\dots,K-1\}$.

\newcommand{\nosemic}{\renewcommand{\@endalgocfline}{\relax}}
\newcommand{\dosemic}{\renewcommand{\@endalgocfline}{\algocf@endline}}
\newcommand{\pushline}{\Indp}
\newcommand{\popline}{\Indm\dosemic}
\let\oldnl\nl
\newcommand{\nonl}{\renewcommand{\nl}{\let\nl\oldnl}}

\begin{algorithm}
\SetAlgoLined \LinesNumbered
{
\caption{Optimal Scheduling Policy}
\KwIn 
{$\left\{ {E_j},{{\beta _j}} \right\}_{j = 1}^U$
} 
\KwOut 
{${\cal O^\star, D^\star, P^\star,} L^\star $
} 
Sort ${E_1}{\beta _1} \ge {E_2}{\beta _2} \ge \dots \ge {E_U}{\beta _U}$.

\For{L=1:N-1}
{
{\nonl{-- \emph{\textbf{First Part}: Find candidate scheduling groups}}}

Set $q\leftarrow 1$.

\For{$1 \le q_1+q_2\le U$, $0 \le q_2-q_1\le M-1$}
{

${\cal O}_q\leftarrow\left\{ { q_1,q_1+1,\dots,q_1+q_2-1} \right\}$.\\

{ Find $\{({p_j^{\rm{tr},\star}},{p_j^{\rm{dt},\star}})\}_{j\in \mathcal{O}_q}$ and $\Omega^\star_q(\mathcal{O}_q,L)$ from (\ref{eq30}), (\ref{eq31}) and (\ref{eq32}).}\\

${\cal P}_q \leftarrow \{({p_j^{\rm{tr},\star}},{p_j^{\rm{dt},\star}})\}_{j\in \mathcal{O}_q}$. \\

$q\leftarrow q +1$.
}

{\nonl{-- \emph{\textbf{Second Part}: Solve the binary integer programming}}}

Construct $\mathbf{c}$ and $\mathbf{S}$ with (\ref{eq35}) and (\ref{eq36}).

Solve the LP (\ref{eq34}) with relaxing $\mathbf{x}\in[0,1]^{C\times 1}$ and let $\mathbf{x}^\star$ be its optimal solution.

Find the index set $\mathcal{Q} = \{q | [\mathbf{x}^\star]_q=1 \}$. 

Compute \[{D_q} = \frac{{{{\left( {\Omega _q^\star\left( {{{\cal O}_q}},L \right)} \right)}^{ - 1}}}}{{\sum\nolimits_{i \in {\cal Q}} {{{\left( {\Omega _i^\star\left( {{{\cal O}_i}},L \right)} \right)}^{ - 1}}} }},~\forall q \in {\cal Q},\] 
and 
\[\mathsf{SE}_{(L)}^ \star = \frac{{1 - {\frac{L}{N}} }}{\eta }\frac{1}{{\sum\nolimits_{i \in \mathcal{Q}} {{{\left( {\Omega _i^ \star \left( {{\mathcal{O}_i}},L \right)} \right)}^{ - 1}}} }}.\]

Store $\mathcal{O}^\star_{(L)} \leftarrow \{\mathcal{O}_q\}_{q\in\mathcal{Q}}$, $\mathcal{P}^\star_{(L)} \leftarrow \{\mathcal{P}_q\}_{q\in\mathcal{Q}}$, and $\mathcal{D}^\star_{(L)} \leftarrow\{D_q\}_{q\in\mathcal{Q}}$.
}
Find the optimal training length, $L^\star = \mathop{\arg\max}\limits_{1\le L<N} \mathsf{SE}_{(L)}^\star$. 

Return $\mathcal{O}^\star \leftarrow \mathcal{O}^\star_{(L^\star)}$, $\mathcal{D}^\star \leftarrow \mathcal{D}^\star_{(L^\star)}$ and $\mathcal{P}^\star \leftarrow \mathcal{P}^\star_{(L^\star)}$.

}
\end{algorithm}

\subsubsection{Optimal Scheduling Group $\mathcal{O}^\star$}
Even though the optimal transmit energy allocation strategy for given users are derived in (\ref{eq30}), the size of the search space is too large to be exhaustively searched. To reduce the search space, we need to find implicit properties for the optimal scheduling groups.

From (\ref{eq23}), the objective function of (P-eq) is given by
$$\mathsf{SE} = \frac{1-{\frac{L}{N}}}{\eta} \frac{1}{\sum\nolimits_{q=1}^Q(\Omega^\star_q(\mathcal{O}^\star_q,L))^{-1}},$$
where $\{\mathcal{O}^\star_q\}_{q=1}^Q$ is the sets of the optimal scheduling groups. Suppose that the cardinality of each of the optimal scheduling groups is given, i.e., $O_q = |\mathcal{O}^\star_q|$. Then, the problem (P-eq) is reduced to the following cardinality-constrained problem for each possible value of $L$:
\begin{subequations}\label{eq33}
\begin{align}
\text{(P-B)}~~& \underset{{{{\cal O}_1},\dots,{{\cal O}_Q}}}{\text{minimize}}
& & \sum\nolimits_{q=1}^Q (\Omega_q^\star (\mathcal{O}_q,L))^{-1},\\
& \text{subject to}
& & |{{\cal O}_q}| = {O_q},~\forall q,\text{ and (\ref{eq19}d)}.
\end{align}
\end{subequations}
To obtain the optimal scheduling groups, the following theorem is quite helpful.

\begin{theorem}
Denote $\mathcal{O}_q^\star$, $\forall q$, as the optimal solution of (P-B) at a given value of $L$. Then the optimal solution has the following properties:
\begin{itemize}
\item [1)] For $q$ with $O_q = 2$, if $\mathcal{O}^\star_q = \{K_1, K_2\}$, then there is no $K_3\in\mathcal{U}$ such that 
$E_{K_1}\beta_{K_1}> E_{K_3}\beta_{K_3}> E_{K_2}\beta_{K_2}$. 

\item [2)] For $q$ with $O_q \ge 3$, if $\{K_1, K_2\}\subset\mathcal{O}^\star_q$ and there exists $K_3\in\mathcal{U}$ such that $E_{K_1}\beta_{K_1}> E_{K_3}\beta_{K_3}> E_{K_2}\beta_{K_2}$, then $K_3\in\mathcal{O}^\star_q$.
\end{itemize}
\end{theorem}
\begin{IEEEproof}
See Appendix B.
\end{IEEEproof}

From Theorem 2, it is shown that the optimal uplink scheduling policy is to select users having similar product values of the average allowed energy level $E_j $ and the path-loss $\beta_j$ and it significantly reduces the search space. More detailed discussions on the search space will be given in Sec. III-D.

Although Theorem 2 indicates some useful implicit properties for the optimal scheduling groups, it does not provide the exact solution explicitly, and we still need to find the optimal scheduling groups among the reduced search space. Fortunately, it can be transformed into a binary integer programming (BIP) with the following generic form:
\begin{subequations}\label{eq34}
\begin{align}
(\text{P-C})~~&\mathop {{\text{minimize }}}\limits_{{{\bf{x}}}} J(\mathbf{x}) = {\mathbf{c}^T\mathbf{x}},\\
&{\text {subject to }}~{{\bf{S}}}{{\bf{x}}} = \mathbf{b},~{{{\mathbf{x}}}} \in \left\{ {0,1} \right\}^{C\times 1},
\end{align}
\end{subequations}
where $\mathbf{S}=\left[ s_{uq} \right]$ is the $U \times C$ state matrix, 
\begin{equation}\label{eq35}
{s_{uq}} = \left\{ {\begin{array}{*{20}{l}}
1,&{u \in \mathcal{O}_q},\\
0,&\text{otherwise},
\end{array}} \right.
\end{equation}
$\mathbf{c}$ is the $C\times 1$ cost vector given by 
\begin{align}\label{eq36}
{\bf{c}} = {\left[ {\frac{1}{{\Omega _1^ \star ({{\cal O}_1},L)}},\frac{1}{{\Omega _2^ \star ({{\cal O}_2},L)}}, \ldots ,\frac{1}{{\Omega _C^ \star ({{\cal O}_C},L)}}} \right]^T},
\end{align}
${\bf{b}} = \mathbf{1}_{C\times 1}$ is the $C\times 1$ all-one vector, 
$C = \left|\left\{\mathcal{O}_q \left| 1\le|\mathcal{O}_q|\le M, \mathcal{O}_q\subset \mathcal{U}\right.\right\}\right|$ denotes the number of candidate scheduling groups, and $\Omega_i^\star(\mathcal{O}_i,L)$ denotes the optimal common rate at given $\mathcal{O}_i$ and $L$, defined in (\ref{eq32}). The optimizing variable $\mathbf{x}$ informs which candidate scheduling groups are selected, i.e., if $x_q =1$, the corresponding candidate scheduling group $\mathcal{O}_q$ is selected as one of the optimal scheduling groups. Such a BIP has been widely researched in literature and a variety of efficient algorithms are summarized in \cite{SchrijverLP}. Unfortunately, finding the optimal solution in a BIP is known as NP-hard in general. However, due to the special structure of our BIP, it will be shown that a linear programming (LP) relaxation using $\mathbf{x}\in[0,1]^{C\times 1}$ does not affect the optimality. To show this, we introduce the following definition and two lemmas and then conclude the optimality of the proposed algorithm.

\begin{definition}
A matrix $\mathbf{A}$ is \emph{totally unimodular} if every square sub-matrix of $\mathbf{A}$ has a determinant of $0$, $-1$, or $1$.
\end{definition}

\begin{lemma} [\cite{SchrijverLP}, Example 7]
If every column of a binary matrix $\mathbf{A}$ has consecutive ones only without being interrupted by $0$s, then $\mathbf{A}$ is totally unimodular.
\end{lemma}

\begin{lemma} [\cite{SchrijverLP}, Theorem 19.1] 
If $\mathbf{S} $ is totally unimodular and $\mathbf{b}$ is an integer vector, then the polytope described by $\mathbf{Sx}=\mathbf{b}$, $\mathbf{x}\in[0,1]^{C\times 1}$, has integer vertices only.
\end{lemma}

Now, we are ready to state the optimality of the proposed algorithm using the LP relaxation in (\ref{eq34}).
\begin{theorem}
The optimal solution of the BIP in (\ref{eq34}) is identical to the solution obtained by using the LP relaxation on (\ref{eq34}). 
\end{theorem}
\begin{IEEEproof}
From the properties of Theorem 2, every column of the matrix $\mathbf{S}$ has consecutive ones only, which implies that $\mathbf{S}$ is totally unimodular from Lemma 4. Using Lemma 5, the feasible region of (\ref{eq34}) is a polytope with integer vertices only, which guarantees that the solution obtained by using the LP relaxation on (\ref{eq34}) does not affect the optimality. 
\end{IEEEproof}

The proposed algorithm for obtaining the optimal static uplink scheduling policy is outlined in Algorithm 1. 
The proposed optimal algorithm obtains $\mathcal{O}^\star_{(L)},\mathcal{D}^\star_{(L)},\mathcal{P}^\star_{(L)}$ and corresponding $\mathsf{SE}^\star_{(L)}$ for each $L$, and then find the optimal $L^\star$ maximizing $\mathsf{SE}^\star_{(L)}$. The algorithm for obtaining $\mathcal{O}^\star_{(L)},\mathcal{D}^\star_{(L)},\mathcal{P}^\star_{(L)}$ is composed of the two parts. The first part finds the candidate scheduling groups, denoted as $\{\mathcal{O}_{q}\}_{q=1}^{C}$, and their corresponding common rate by using (\ref{eq32}) obtained by using the optimal energy allocations in Theorem 1. Then, the second part finds the optimal combination of the selected scheduling groups that maximizes the spectral efficiency by applying the LP relaxation by virtue of Theorems 2 and 3.

\emph{Example:} Here, we explain a toy example. Suppose that the network has $U=4$ users, $\{1,2,3,4\}$, and $M =2$ and $\mathcal{T}_{\rm{th}} = 10$Kbits, $W = 1$KHz, $F=16$, $N=8$, $L=4$, and $\eta = 1$. The first part returns the candidate scheduling groups as ${{\cal O}_1} = \left\{ 1 \right\},{{\cal O}_2} = \left\{ 2 \right\},{{\cal O}_3} = \left\{ 3 \right\},{{\cal O}_4} = \left\{ 4 \right\},{{\cal O}_5} = \left\{ {1,2} \right\},{{\cal O}_6} = \left\{ {2,3} \right\}{{\cal O}_7} = \left\{ {3,4} \right\}$ and suppose that the corresponding common rate is determined as ${\Omega _1} = 11,{\Omega _2} = 10,{\Omega _3} = 5,{\Omega _4} = 3,{\Omega _5} = 9,{\Omega _6} = 4,{\Omega _7} = 2$, respectively. Then, the cost vector and the state matrix are respectively set as 
\[{\bf{c}} = {\left[ 
{\frac{1}{{11}}},{\frac{1}{{10}}},{\frac{1}{5}},{\frac{1}{3}},{\frac{1}{9}},{\frac{1}{4}},{\frac{1}{2}},{\frac{1}{3}},1
\right]^T}\]
and
\[{\bf{S}} = \left[ {\begin{array}{*{20}{c}}
1&0&0&0&1&0&0\\
0&1&0&0&1&1&0\\
0&0&1&0&0&1&1\\
0&0&0&1&0&0&1
\end{array}} \right].\]
To satisfy the constraints (\ref{eq34}b), there exist five feasible solutions: $\mathbf{x}_{1} = \left[ {1,1,1,1,0,0,0} \right]^T$, $\mathbf{x}_{2} = \left[ {1,1,0,0,0,0,1} \right]^T$, $\mathbf{x}_{3} = \left[ {1,0,0,1,0,1,0} \right]^T$, ${\bf{x}}_4 = \left[ {0,0,1,1,1,0,0} \right],{\bf{x}}_5 = \left[ {0,0,0,0,1,0,1} \right]$. Since $J\left( {{{\bf{x}}_1}} \right) = 0.7242 > J\left( {{{\bf{x}}_2}} \right) = 0.6909 > J\left( {{{\bf{x}}_3}} \right) = 0.6742 > J\left( {{{\bf{x}}_4}} \right) = 0.6444 > J\left( {{{\bf{x}}_5}} \right) = 0.6111$, $\mathcal{O}^\star_1 = \mathcal{O}_5=\{1,2\}$ and $\mathcal{O}^\star_2 = \mathcal{O}_7=\{3,4\}$ are selected as the optimal uplink scheduling groups. And $D_1 = 0.1818$ and $D_2 = 0.8182$ are the optimal scheduling portions and $\mathsf{SE} = 9/11$ bps/Hz. Then, the latency is $\lceil 10/(9/11) \rceil = 13$ frames. Note that among $13\times 16=208$ sub-frames $208\times 0.1818 \approx 38$ sub-frames are allocated to $\mathcal{O}_1^\star$ and the remained sub-frames are allocated to $\mathcal{O}_2^\star$.

\begin{table*}[ht!]
\centering
\caption{Simulation Results using the ZF Receiver with $M=64$ and $U=100$}
\scalebox{0.93}{
\begin{tabular}{?m{0.3cm}?m{2cm}?m{1.17cm  }|m{1.17cm  }|m{1.17cm  }|m{1.17cm  }|m{1.17cm  }|m{1.17cm  }|m{1.17cm  }|m{1.17cm  }|m{1.17cm  }|m{1.17cm  }?}
\specialrule{.1em}{.05em}{.05em} 
{}&$E$ (dB) & 50 & 60 & 70 & 80 & 90 & 100 & 110 & 120 & 130 & 140 \\
\hline\hline
{RE}&Latency (sec) & 6.320e+2 & 6.869 & 1.514e-1 & 2.483e-2 & 1.272e-2 & 8.898e-3 & 6.983e-3 & 5.771e-3 & 4.924e-3 & 4.296e-3
\\
\specialrule{.1em}{.05em}{.05em} 
{RO}&Latency (sec)  &2.386e+2 & 2.567 & 4.524e-2 & 3.698e-3 & 1.227e-3 & 6.896e-4 & 4.698e-4 & 3.489e-4 & 2.774e-4 & 2.303e-4
\\
\specialrule{.1em}{.05em}{.05em} 
\multirow{6}{*}{\rotatebox{90}{{\multirow{2}{*}{Proposed}}}}&Latency (sec)  &8.617e+1& 9.286e-1& 1.742e-2& 2.075e-3& 8.762e-4 &5.458e-4& 3.990e-4& 3.126e-4& 2.535e-4& 2.133e-4\\
\cline{2-12}
{}&$L^\star$ & 20 & 20 & 20 & 25 & 34 & 34 & 34 & 50 & 50 & 50 \\
\cline{2-12}
{}&$\left[|\mathcal{O}^\star_1|,...,|\mathcal{O}^\star_Q|\right]$ & [8 11 13 14 16 18 20] & [13 16 16 17 18 20] & [20 20 20 20 20] & [25 25 25 25] & [32 34 34] & [32 34 34] & [32 34 34] & [50 50] & [50 50] & [50 50] \\
\cline{2-12}
{}&$\mathsf{SE}^\star$ (bps/Hz) & 1.160e-5&1.077e-3&5.740e-2&4.819e-1&1.141&1.832&2.506&3.199&3.945&4.689 \\
\specialrule{.1em}{.05em}{.05em} 
\end{tabular}
}
\end{table*}
\begin{table*}[ht!]
\centering
\caption{Simulation Results using the ZF Receiver with  $E=70\mathrm{dB}$ and $U=100$}
\scalebox{0.93}{
\begin{tabular}{?m{0.3cm}?m{2cm}?m{1.17cm  }|m{1.17cm  }|m{1.17cm  }|m{1.17cm  }|m{1.17cm  }|m{1.17cm  }|m{1.17cm  }|m{1.17cm  }|m{1.17cm  }|m{1.17cm  }?}
\specialrule{.1em}{.05em}{.05em} 
{}&$M$ & 32 & 64 & 128 & 256 & 512 & 1024 & 2048 & 4096 & 8192 & 16384 \\
\hline\hline
{RE}&Latency (sec) & 2.663e-1 & 1.514e-1 & 9.166e-2 & 5.968e-2 & 4.174e-2 & 3.112e-2 & 2.445e-2 & 2.001e-2 & 1.690e-2 & 1.463e-2
\\
\specialrule{.1em}{.05em}{.05em} 
{RO}&Latency (sec) & 1.330e-1 & 4.524e-2 & 1.604e-2 & 7.318e-3 & 3.978e-3 & 2.188e-3 & 1.310e-3 & 8.570e-4 & 6.020e-4 & 4.479e-4
\\
\specialrule{.1em}{.05em}{.05em} 
\multirow{6}{*}{\rotatebox{90}{{\multirow{2}{*}{Proposed}}}}&Latency (sec) & 4.541e-2 & 1.742e-2 & 7.821e-3 & 4.058e-3 & 2.449e-3 & 1.646e-3 & 1.146e-3 & 8.384e-4 & 6.020e-4 & 4.479e-4 \\
\cline{2-12}
{}&$L^\star$ & 13 & 20 & 26 & 34 & 34 & 45 & 38 & 32 & 29 & 25 \\
\cline{2-12}
{}&$\left[|\mathcal{O}^\star_1|,...,|\mathcal{O}^\star_Q|\right]$& [12 13 13 13 12 12 12 13] & [20 20 20 20 20] & [22 26 26 26] & [32 34 34] & [32 34 34] & [45 55] & [38 62] & [32 68] & [100] & [100] \\
\cline{2-12}
{}&$\mathsf{SE}^\star$ (bps/Hz) &2.202e-2 & 5.740e-2 & 1.279e-1 & 2.464e-1 & 4.083e-1 & 6.075e-1 & 8.723e-1 & 1.193 & 1.661 & 2.233\\
\specialrule{.1em}{.05em}{.05em} 
\end{tabular}}
\end{table*}
\begin{table*}[ht!]
\centering
\caption{Simulation Results using the ZF Receiver with $M=64$ and $E=70\mathrm{dB}$}
\scalebox{0.93}{
\begin{tabular}{?m{0.3cm}?m{2cm}?m{1.17cm  }|m{1.17cm  }|m{1.17cm  }|m{1.17cm  }|m{1.17cm  }|m{1.17cm  }|m{1.17cm  }|m{1.17cm  }|m{1.17cm  }|m{1.17cm  }?}
\specialrule{.1em}{.05em}{.05em} 
{}&$U$ & 100 & 200 & 300 & 400 & 500 & 600 & 700 & 800 & 900 & 1000 \\
\hline\hline
{RE}&Latency (sec) &1.514e-1 & 2.988e-1 & 4.462e-1 & 5.936e-1 & 7.410e-1 & 8.884e-1 & 1.036 & 1.183 & 1.331 & 1.478
\\
\specialrule{.1em}{.05em}{.05em} 
{RO}&Latency (sec) &4.524e-2 & 9.238e-2 & 1.330e-1 & 1.701e-1 & 2.229e-1 & 2.587e-1 & 3.020e-1 & 3.523e-1 & 3.873e-1 & 4.379e-1
\\
\specialrule{.1em}{.05em}{.05em} 
\multirow{14}{*}{\rotatebox{90}{{\multirow{2}{*}{Proposed}}}}&Latency (sec) &1.742e-2 & 2.708e-2 & 3.668e-2 & 4.637e-2 & 5.603e-2 & 6.572e-2 & 7.531e-2 & 8.501e-2 & 9.471e-2 & 1.043e-1\\
\cline{2-12}
{}&$L^\star$ & 20 & 25 & 25 & 27 & 27 & 27 & 28 & 28 & 28 & 28 \\
\cline{2-12}
{}&$\left[|\mathcal{O}^\star_1|,...,|\mathcal{O}^\star_Q|\right]$& [20 20 20 20 20] & [25 25 25 25 25 25 25 25] & [25 25 25 25 25 25 25 25] & [22 27 27 27 27 27 27 27 27 27 27 27 27 27 27] & [14 27 27 27 27 27 27 27 27 27 27 27 27 27 27 27 27 27 27] & [6 27 27 27 27 27 27 27 27 27 27 27 27 27 27 27 27 27 27 27 27 27 27] & [28 28 28 28 28 28 28 28 28 28 28 28 28 28 28 28 28 28 28 28 28 28 28 28 28] & [16 28 28 28 28 28 28 28 28 28 28 28 28 28 28 28 28 28 28 28 28 28 28 28 28 28 28 28 28] & [4 28 28 28 28 28 28 28 28 28 28 28 28 28 28 28 28 28 28 28 28 28 28 28 28 28 28 28 28 28 28 28 28] & [20 28 28 28 28 28 28 28 28 28 28 28 28 28 28 28 28 28 28 28 28 28 28 28 28 28 28 28 28 28 28 28 28 28 28 28] \\
\cline{2-12}
{}&$\mathsf{SE}^\star$ (bps/Hz) & 5.740e-2 & 3.692e-2 & 2.726e-2 & 2.157e-2 & 1.785e-2 & 1.522e-2 & 1.328e-2 & 1.176e-2 & 1.056e-2 & 9.588e-3 \\
\specialrule{.1em}{.05em}{.05em} 
\end{tabular}}
\end{table*}

\subsection{Computational Complexity Analysis}
\DeclareRobustCommand{\stirling}{\genfrac\{\}{0pt}{}}

Define 
\begin{equation}\label{eq37}
\mathcal{F}(U,M) = \bigcup\nolimits_{Q=\lfloor U/M \rfloor}^U {\cal F}(U,{M},Q)
\end{equation}
 as the whole search space for finding the optimal scheduling groups in (\ref{eq19}) without Theorem 2, where ${\cal F}(U,{M},Q)$ is the collection of $Q$-ary partitions of $\mathcal{U}$ with at most $M$ elements, i.e., each scheduling group size is no greater than the number of antennas, given by
\begin{equation}\label{eq38}
\begin{split}
&{\cal F}(U,{M},Q) = \left\{ {({{\cal O}_1},\dots,{{\cal O}_Q})\left| \begin{array}{l}
1 \le |{{\cal O}_q}| \le {M},~{{\cal O}_p} \cap {{\cal O}_q} = \emptyset ,\\
{{\cal O}_1} \cup {{\cal O}_Q} = \mathcal{U} 
\end{array} \right.} \right\}.\end{split}
\end{equation}
Note that $\mathcal{F}(U,M,Q)$ is well-defined only when $U\le Q M$ and $\left|\mathcal{F}(U,U,Q)\right| = \stirling{U}{Q}$, where $\stirling{U}{Q}$ denotes the Stirling number of the second kind. For a fixed $Q$, $\stirling{U}{Q}\sim Q^U/Q!$ increases exponentially with $U$. Thus, the whole search space is given as $\left|\mathcal{F}(U,M)\right|>Q^U/Q! $ for a large $U$. 

Now, define $\mathcal{F}^{\rm{r}}(U,M)$ as the reduced search space for finding the optimal scheduling groups in (\ref{eq19}) with the aid of Theorem 2. Then, we can show that the cardinality of $\mathcal{F}^{\rm{r}}(U,M)$ can be represented as the following recursive formula:
\begin{equation}\label{eq39}
\left|\mathcal{F}^{\rm{r}}(U,M)\right|= \sum\nolimits_{k=1}^{M}\left|\mathcal{F}^{\rm{r}}(U-k,M)\right|,
\end{equation}
 which is known as the generalized Fibonacci number \cite{DresdenBinet}. With help of the Binet's formula \cite{DresdenBinet}, we arrive at 
$$ \left|\mathcal{F}^{\rm{r}}(U,M)\right| = \left\lceil \frac{w-1}{2+(M+1)(w-2)}2^{U-1}\right\rfloor,
$$
where $w$ is the unique positive root of $x^{M}-x^{M-1}-\dots-1=0$. After some algebraic manipulations, $w\to 2$ and $(M+1)(w-2)\to 0$ as $M\to \infty$ so that $\left|\mathcal{F}^{\rm{r}}(U,M)\right| \to 2^{U-2}$, which implies that the reduced search space still increases exponentially with the number of total users, $U$. However, combined with Theorem 3, the following dramatical complexity reduction can be obtained. 
\begin{itemize}
\item[1)] The reduction gain of Theorem 2 itself also increases exponentially with $U$. In fact, the reduction gain is at least $(Q/2)^U/Q!$ for a large $U$. 
\item[2)] Without Theorem 2, the number of candidate scheduling groups in (\ref{eq34}) is $C = \sum_{k=1}^{M}{U\choose k}$, (for $M = U$, $C= 2^U-1$), which increases exponentially with $U$. However, due to Theorem 2, it reduces into $C = \sum_{k=1}^{M}(U-k+1) = \frac{1}{2}{{{M}}}\left( {2U - M + 1} \right)$, (for $M = U$, $C= \frac{1}{2}U(U+1)$), which increases only squarely with $U$. 
\item[3)] As shown in Theorem 3, applying the LP relaxation on (\ref{eq34}) still provides the optimal solution of the original BIP (\ref{eq34}), which can dramatically reduce the exponential-time complexity into a polynomial-time complexity. 
\end{itemize}

Now, we are ready to quantify the computational complexity of Algorithm 1. 
The computational complexity of Algorithm 1 consists of the following three parts, namely 1) the sorting operation (line 1), 2) the optimal energy allocation (lines 3-9) and 3) solving the relaxed LP (lines 10-15). The worst-case computational complexity for sorting $U$ samples is $O(U\log U)$. Since the optimal energy allocation requires $\frac{1}{2}M\left( {2U - M + 1} \right)$ iterations, the worst-case computational complexity of the second part is $O({M}U)$. Finally, the worst-case computational complexity of the LP is $O\left((MU)^{3.5}\right)$ by using the Karmarkar's algorithm \cite{KarmarkarLP}. Thus, the total worst-case computational complexity for the proposed algorithm is $O(U\log U + N{M}U +N({M}U)^{3.5})=O(N({M}U)^{3.5})$.

\section{Numerical Results}
In this section, we present some numerical results to verify the superiority of the proposed uplink scheduling policy. One frame is set to occupy $10$MHz and $1$ms in the frequency and time domains and consists of $F=80$ sub-frames with $125$KHz and $1$ms. The number of symbols in each sub-frame is set to $N = 100$ by assuming $\eta = 1.25$ (25\% CP overhead). 
There are $U=300$ users each requesting $\mathcal{T}_{\rm{th}}=10$Kbits date volume.
We use the pathloss model $\beta_{j} = G_0d_j^{-\alpha}$, where $G_0= 0.1$, $\alpha=4$ and $d_j$ is given by 
\[d_j = R_{\min} +\frac{(R_{\max}-R_{\min})j}{U},\]
with $R_{\min} = 10, R_{\max} =100$. This pathloss model reflects the BS located at the origin and the users are located uniformly along the line $[R_{\min},R_{\max}]$. All of users have the same transmit energy constraint, $E_j=E$ for $\forall j$. According to the simulation setting, the received signal energy of the worst-case user at the BS is $0$dB when $E = 70$dB energy is equally spread over the symbols in  a sub-frame.\footnote{Note that by assuming $-174$dBm/Hz for the noise spectral density, $E=90$dB means only $-3$dBm ($0.5$mW) per sub-frame in this simulation setting.} 

\begin{figure}
\centering
\includegraphics[width=0.5\columnwidth]{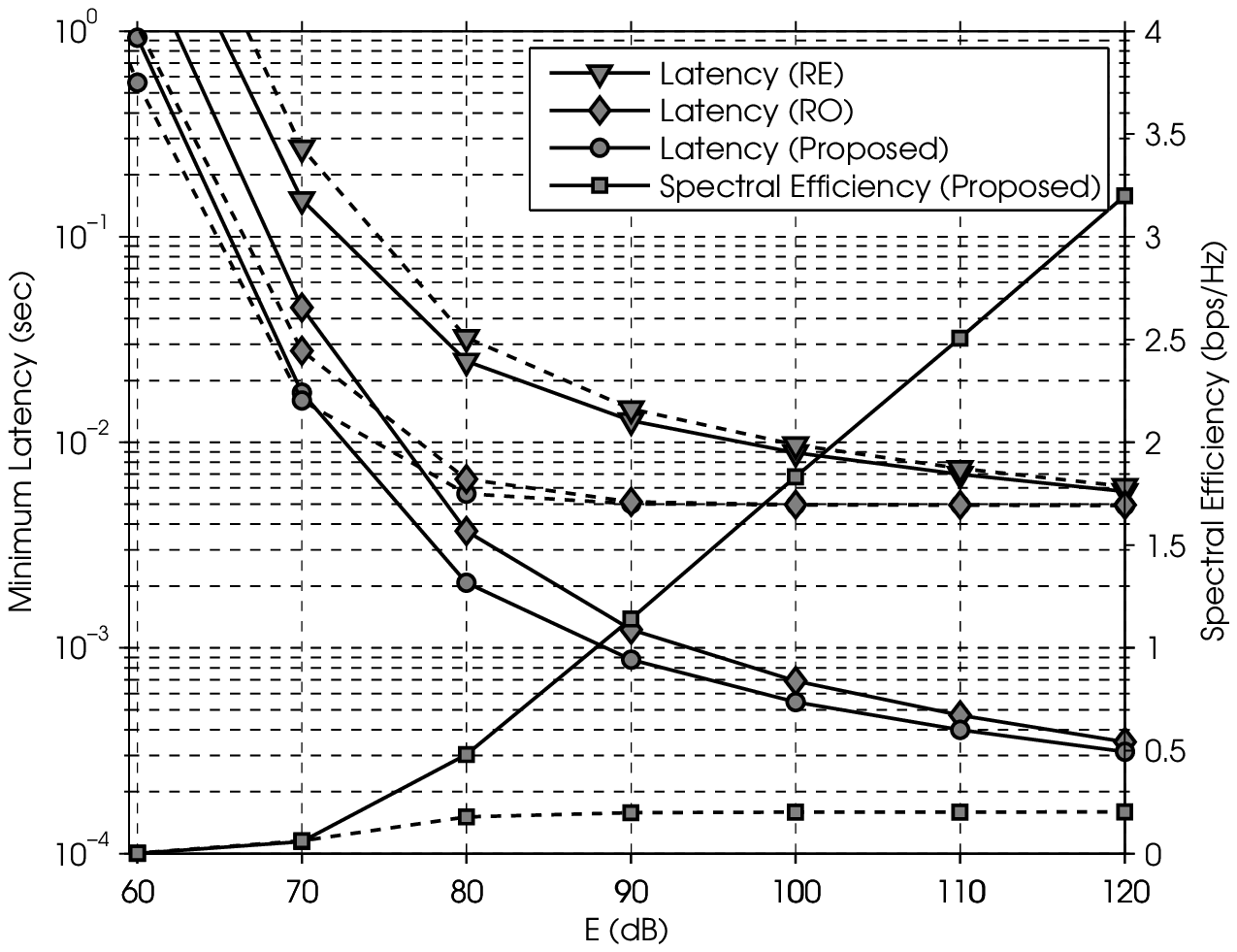}
\caption{Two-sided plot for latency and spectral efficiency as a function of $E$ when $M=64$ and $U=100$. The solid line represents the results using the ZF receiver and the dashed line represents the results using the MRC receiver.}
\end{figure}

\begin{figure}
\centering
\includegraphics[width=0.5\columnwidth]{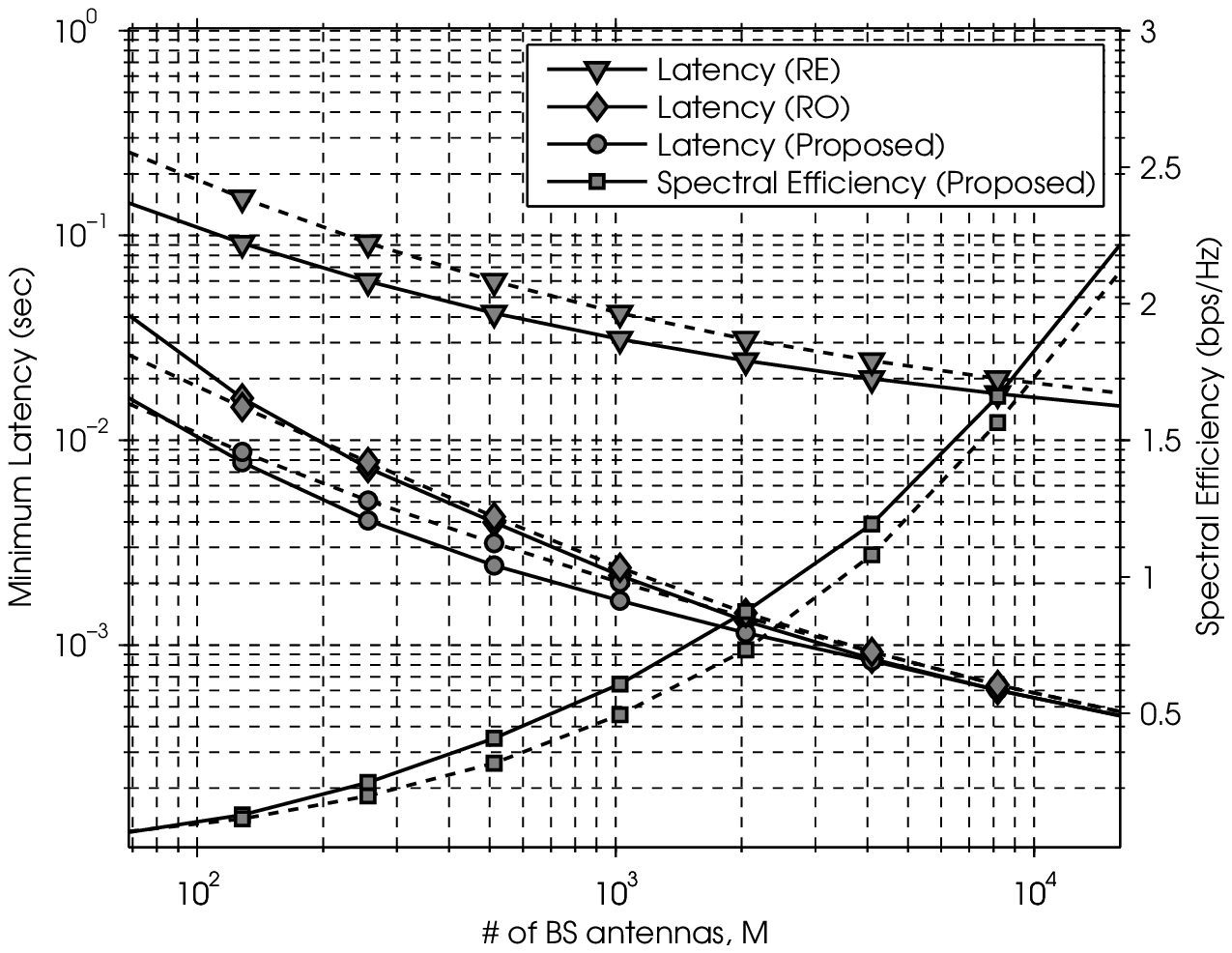}
\caption{Two-sided plot for latency and spectral efficiency as a function of $M$ when $E=70$dB and $U=100$. The solid line represents the results using the ZF receiver and the dashed line represents the results using the MRC receiver.}
\end{figure}

\begin{figure}
\centering
\includegraphics[width=0.5\columnwidth]{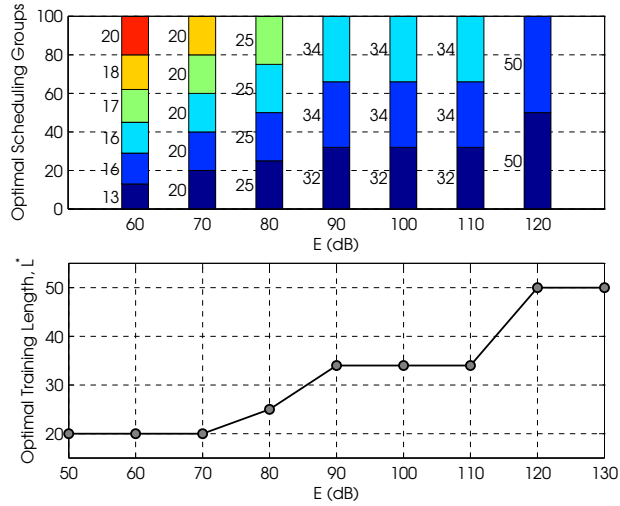}
\caption{Optimal uplink scheduling parameters as a function of $E$ when $M=64$ and $U=100$.}
\end{figure}
\begin{figure}
\centering
\includegraphics[width=0.5\columnwidth]{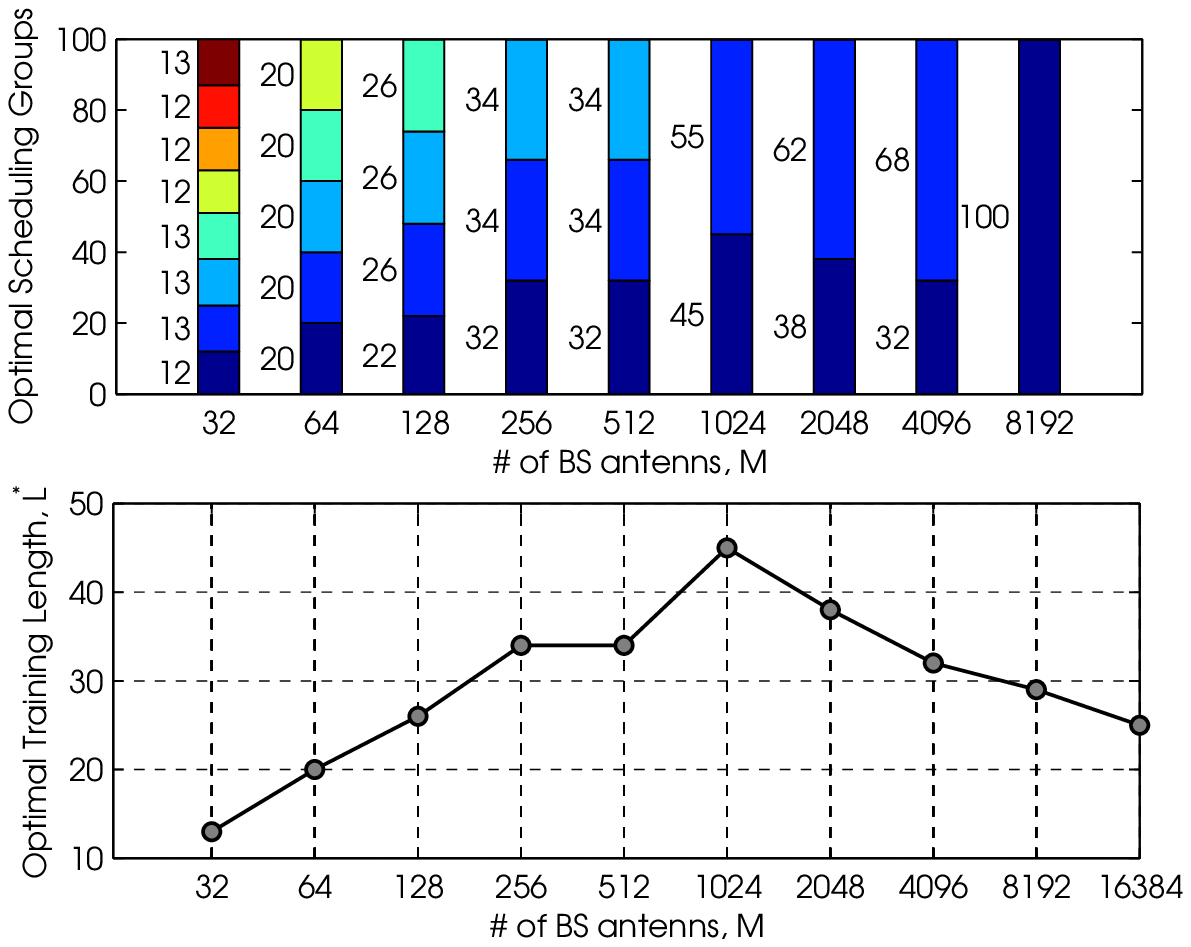}
\caption{Optimal uplink scheduling parameters as a function of $M$ when $E=70$dB and $U=100$.}
\end{figure}

The following three schemes are compared and simulation results using the ZF receiver are summarized in Tables III-V:
\begin{itemize}
\item[1)] (Random-Equal (RE)) $K$ users are randomly selected and the transmit energy is equally used during the training and data transmission phases. The training length, $L$, and $K$ are exhaustively searched.
\item[2)] (Random-Optimal (RO)) $K$ users are randomly selected and the transmit energy is optimized by using Theorem 1. The training length, $L$, and $K$ are exhaustively searched.
\item[3)] (Proposed) the optimal uplink scheduling policy in Algorithm 1 is used.
\end{itemize}

Fig. 4 depicts the latency and spectral efficiency of the three schemes as a function of $E$ when $M=64$ and $U=100$. The solid line represents the results using the  ZF receiver and the dashed line represents the results using the MRC receiver. When the ZF receiver is employed, it is observed that at $E=80$dB ($-13$dBm per sub-frame), the proposed uplink scheduling policy provides about $12.0$ or $1.78$ times smaller latency over the Random-Equal or the Random-Optimal scheme. The major gain comes from the optimal energy allocation. When the MRC receiver is employed, it is observed that at $E=80$dB the proposed uplink scheduling policy provides about $5.74$ or $1.18$ times smaller latency over the Random-Equal or the Random-Optimal scheme. Similarly as in the ZF case, the major gain comes from the optimal energy allocation. The difference is that the common  spectral efficiency of the ZF receiver increases logarithmically with $E$, while that using the MRC receiver is saturated at high $E$ due to the uncanceled interference. So, when high $E$ is available, the ZF receiver clearly outperforms the MRC receiver. The gain of the ZF receiver over the MRC receiver is $2.71$ and it becomes $9.08$ at $E=80$dB.

Fig. 5 depicts the latency and spectral efficiency of the three schemes as a function of $M$ when $E=70$dB and $U=100$. When the ZF receiver is employed, it is observed that at $M=256$, the proposed uplink scheduling policy provides about $14.7$ or $1.80$ times smaller latency over the Random-Equal or the Random-Optimal scheme. The gain of the proposed one over the Random-Optimal scheme diminishes when the number of BS antennas becomes high because all users can be scheduled with sharing the same resource. Similar trends can be observed when the MRC receiver is employed and the latency and common  spectral efficiency of the two receivers become identical. 


\begin{figure}
\centering
\subfigure[\smaller $E=70$dB, $M=64$ and $U=100$]{\includegraphics[width=0.5\columnwidth]{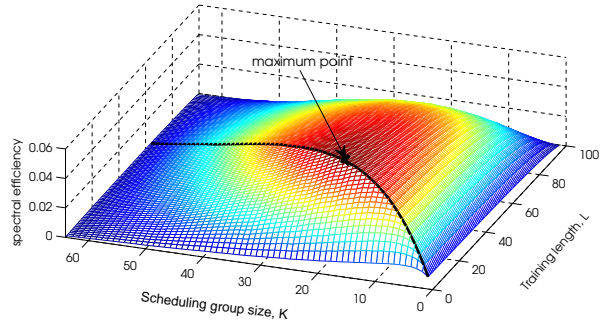}}
\subfigure[\smaller $E=60$dB, $M=64$ and $U=100$]{\includegraphics[width=0.5\columnwidth]{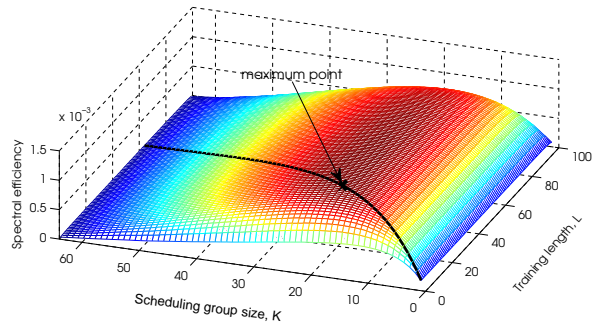}}
\subfigure[\smaller $E=70$dB, $M=2048$ and $U=100$]{\includegraphics[width=0.5\columnwidth]{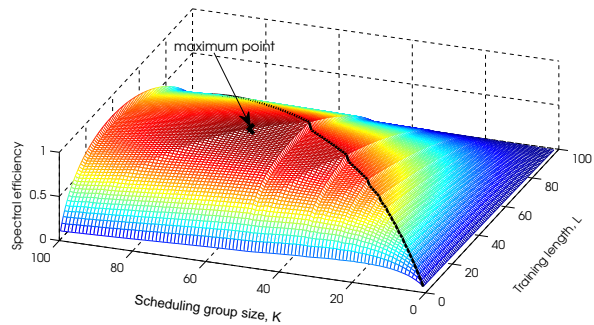}}
\caption{3D plot of the common spectral efficiency as a function of $L$ and $K$. The line indicates where $L=K$ and the marker $\mathsf{x}$ represents the maximum point.}
\end{figure}

Now, discussions on the behavior of the proposed uplink scheduling policy are provided. Fig. 6 depicts the optimal scheduling groups, $\left[|\mathcal{O}_1^\star|,\dots,|\mathcal{O}_Q^\star|\right]$, and the optimal training length, $L^\star$, as a function of $E$ when $M=64$, $U=100$, and the ZF receiver is employed. Fig. 6 shows that small-size scheduling groups are preferred at low $E$, while large-size ones are preferred at high $E$, because high array gain is required at low $E$. 
In spite of the pathloss difference, the size of each optimal scheduling group is nearly identical. 
Although a longer training period is required in high $E$, it is less than or equal to $L^\star = 50$, which is the half of each sub-frame. 
Over a wide range of $E$, $L^\star$ is larger than or equal to $|\mathcal{O}_q^\star|$ for $\forall q$, which implies that orthogonal pilots can be used for not-so-large number of $M$. Also, small-size scheduling groups are preferred at low $M$, while large-size ones are preferred at high $M$ because higher array gain is available. 

Fig. 7 illustrates the optimal scheduling groups, $\left[|\mathcal{O}_1^\star|,\dots,|\mathcal{O}_Q^\star|\right]$, and the optimal training length, $L^\star$, as a function of $M$ when $E=70$dB, $U=100$, and the ZF receiver is employed. Similar trends are observed as in Fig. 6 for the size of each optimal scheduling group. 
On the other hand, the optimal training length first increases with $M$ but it becomes decreasing if $M$ increases further. Interestingly, $L^\star$ becomes smaller than $|\mathcal{O}_q^\star|$ for large $M$, which implies that non-orthogonal pilots become beneficial. In such a case, very high array gain is available and the optimal uplink scheduling policy provides an efficient non-orthogonal multiple access among users so that a low-latency ultra-reliable communication can be provided.

Suppose that users are partitioned into $\mathcal{O}_q=\{(q-1)K+1,\dots,\min\{qK,U\}\}$ for $q=1,\dots,\lceil U/K \rceil$ and Figs. 8 (a)-(c) visualize the effect of the training length $L$ and the scheduling group size $K$ on the common spectral efficiency.\footnote{In the next section, it will be shown that such scheduling groups  become optimal as the number of total users increases.} Note that the support of $(L,K)$ is partitioned into  two regions separated by the line $L=K$, because non-orthogonal pilots are used in $K>L$, while only orthogonal pilots are used in $K\le L$. In Figs. 8 (a) and (b), it is seen that the maximum points are belongs to the orthogonal region and the choice of the training length does not affect much for low $E$ and not high $M$. However, as can be seen in Fig. 8 (c), the optimal choice of $L$ does matter and it belongs to the non-orthogonal region at high $M$. 

So far, the superiority of the proposed uplink scheduling policy is verified by using some numerical examples, which indicates that the behavior of the optimal uplink scheduling policy changes significantly according to the system parameters $E$ and $M$. In order to provided better insight on this, an asymptotic analysis would be fruitful.

\section{Asymptotic Analysis}
It is interesting to consider the case that the total number of users, $U$, and the number of BS antennas, $M$ are simultaneously large, but $U$ is far larger than $M$.\footnote{In this section, we concentrate on the ZF receiver, but the analysis can be directly extended to the MRC receiver.} 
The product of each user's location (or pathloss $\{\beta_j\}$) and the transmit energy constraint $\{E_j\}$ is considered as independent random process with a common distribution $F(x)$\footnote{It includes any independent point process for the user locations with allowing sufficient energy for any given location-aware independent power compensation policy.} and we denote $\rho = \mathbb{E}[E_j\beta_j]$ as the averages of the transmit and the receive energies, respectively.



The following theorem states the asymptotic behavior of the static uplink scheduling policy and its network latency.

\begin{theorem}
Let $X$ denote a random variable with cdf $F(x)$. Suppose that $H(L,K)=\mathbb{E}[\log_2^{-1}(1+h(X;L,K))]$ exists for $h(x;L,K)$ given as in the bottom of this page. Then, as $U\to\infty$, the followings hold:
\begin{figure*}[b]
\hrulefill
\vspace*{4pt}
{
\begin{equation}\label{eq40}
\begin{split}
&h(x;L,K) =\\& \left\{ {\begin{array}{*{20}{l}}
{\frac{{L\left( {M - K} \right)}}{K}\frac{{KNx + 2L\left( {N - L} \right)\left( {1 - \sqrt {\left( {\frac{K}{{N - L}}x + 1} \right)\left( {\frac{K}{L}x + 1} \right)} } \right)}}{{\left( {K - L} \right)\left( {KNx + 2L\left( {N - L} \right)\left( {1 - \sqrt {\left( {\frac{K}{{N - L}}x + 1} \right)\left( {\frac{K}{L}x + 1} \right)} } \right)} \right) + K{{\left( {N - 2L} \right)}^2}}},}&{{\text{if }}K > L,}\\
{\frac{{\left( {M - K} \right)\left( {\left( {N - L + K} \right)x + 2\left( {N - L} \right)\left( {1 - \sqrt {\left( {\frac{K}{{N - L}}x + 1} \right)\left( {x + 1} \right)} } \right)} \right)}}{{{{\left( {N - L - K} \right)}^2}}},}&{{\text{if }}K \le L,N \ne L + K,}\\
{\frac{{\left( {M - K} \right)}}{{4K}}\frac{{{x^2}}}{{x + 1}},}&{{\text{if }}K \le L,N = L + K.}
\end{array}} \right.
\end{split}
\end{equation}}
\end{figure*}
\begin{itemize}
\item[1)] The asymptotically optimal pilot length, $L^\star$, and the optimal scheduling group size, $K^\star$, are respectively given by
\begin{equation}\label{eq41}
\left( {{L^ \star },{K^ \star }} \right) = \mathop {\arg \min }\limits_{\scriptstyle 1 \le L < N, \atop \scriptstyle 1 \le K \le M} \frac{H(L,K)}{{K\left( {1 - \dfrac{L}{N}} \right)}},
\end{equation}
\item[2)] and the network latency normalized by $U$ converges to 
\begin{equation}\label{eq42}
\frac{\mathcal{D}^\star}{U} \xrightarrow{{a.s.}} {\overline{\mathcal{D}}^\star},
\end{equation} 
where
\begin{align}\label{eq43}
{\overline {\mathcal{D}}^ \star } = \frac{{\eta {\mathcal{T}_{{\rm{th}}}}}}{W} {\frac{H(L^\star,K^\star)}{{{{K}^ \star }\left( {1 - \dfrac{{{{L}^\star }}}{N}} \right)}}}.
\end{align}
\end{itemize} 
\end{theorem}
\begin{IEEEproof}
See Appendix C.
\end{IEEEproof}

Theorem 4 implies that a set of equi-sized scheduling groups become asymptotically optimal as $U\to\infty$. So, it gives a hint to construct an easy way to implement an asymptotically optimal scheduling policy, which is outlined in Algorithm 2. The computational complexity of the proposed asymptotically optimal scheduling policy is just ${O}(U\log U+U) = O(U\log U)$, mainly comes form  the sorting operation.

\begin{algorithm}[t]
\SetAlgoLined \LinesNumbered
{
\caption{Asymptotically Optimal Scheduling Policy}
\KwIn 
{$\left\{ E_j,{\beta _j} \right\}_{j = 1}^U$
} 
\KwOut 
{${\cal O^\star, D^\star, P^\star,} L^\star $
} 

Assume $(L^\star,K^\star)$ is already computed by using (\ref{eq41}) for a known distribution according to the user locations and allowed power compensation polices.

Sort ${E_1}{\beta _1} \ge {E_2}{\beta _2} \ge \dots \ge {E_U}{\beta _U}$.

$Q \leftarrow \lceil U/{K^\star}\rceil$. 

\For{ $q=1:Q$}{
$\mathcal{O}^\star_q \leftarrow \{(q-1)K^\star +1,\dots, \min\{ qK^\star , U\}\}.$

{ Find $\{({p_j^{\rm{tr},\star}},{p_j^{\rm{dt},\star}})\}_{j\in \mathcal{O}^\star_q}$ and $\Omega^\star_q(\mathcal{O}^\star_q,L^\star)$ from (\ref{eq30}), (\ref{eq31}) and (\ref{eq32}).}\\

${\cal P}^\star_q \leftarrow \{({p_j^{\rm{tr},\star}},{p_j^{\rm{dt},\star}})\}_{j\in \mathcal{O}^\star_q}$. \\
}
Compute \[{D^\star_q} = \frac{{{{\left( {\Omega _q^\star\left( {{{\cal O}^\star_q}},L^\star \right)} \right)}^{ - 1}}}}{{\sum\nolimits_{i=1}^{Q} {{{\left( {\Omega _i^\star\left( {{{\cal O}^\star_i}},L^\star \right)} \right)}^{ - 1}}} }},~\forall q=1,\dots,Q.\] 

Return $\mathcal{O}^\star \leftarrow \{\mathcal{O}^\star_q\}$, $\mathcal{D}^\star \leftarrow \{\mathcal{D}^\star_q\}$ and $\mathcal{P}^\star \leftarrow \{\mathcal{P}^\star_q\}$.
}
\end{algorithm}

{Since Theorem 4 is involved with a complicated function $h(x;L,K)$, it is not easy to gain a good insight. For a further insight, we restrict the random variable $E_j\beta_j$ with $\mathrm{Var}(E_j\beta_j)\ll \rho^2 $, which implies that almost all realizations of $E_j\beta_j$ are scaled as $\rho$. In addition, the number of BS antennas is assumed to be large, but is much smaller than $U$. Then, the following four scaling regimes according to the number of BS antennas, $M$, and the average received energy constraint level $\rho$,} can be classified:

\begin{itemize}
\item[i)] $\rho\gg1$ and $ M\ll\log^2\rho$: sufficiently high $\rho$ and not-so-large $M$,
\item[ii)] $\rho\gg1$ and $M\gg\log^2\rho$: sufficiently high $\rho$ and large $M$,
\item[iii)] $\rho\ll 1$ and $ M\ll 1/\rho$: sufficiently low $\rho$ and not-so-large $M$,
\item[iv)] $\rho\ll 1$ and $M\gg 1/\rho$: sufficiently low $\rho$ and large $M$.
\end{itemize}

Then, the following theorem states the asymptotic behavior of the proposed optimal static uplink scheduling policy.

\begin{theorem} Suppose that $\mathrm{Var}(E_j\beta_j)\ll \rho^2$.
As $U\to\infty$ and $M\to\infty$ with $U/M\to\infty$, the followings hold.
\begin{itemize}
\item[1)]
The asymptotically optimal training length and the scheduling group size are respectively given by 
\begin{equation}\label{eq44}
\begin{split}
&\left( {{L^ \star },{K^ \star} } \right)=\left\{ 
\begin{array}{*{20}{l}}
{\left( { \left\lceil {N}/{2}\right\rfloor , \left\lceil {N}/{2} \right\rfloor } \right),}&{{\text{if Regime i),}}}\\
{\left( { \left\lceil {N}/{3} \right\rfloor , \left\lceil \chi^\star \sqrt {MN} +o(\sqrt{M})\right\rfloor  } \right),}&{{\text{if Regime ii),}}}\\
\left( {l, \left\lceil {M}/{2} \right\rfloor } \right),&{\text{if Regimes iii) \& iv)}},
\end{array}\right.
\end{split}
\end{equation}
where $l$ is an arbitrary integer among $1\le l<N$, 
$$
\chi^\star = \sqrt {\dfrac{1}{3}\left( {\dfrac{2}{{\mathsf{W}\left( { - 2{e^{ - 2}}} \right) + 2}} - 1} \right)} \approx 0.2915,
$$
and $\mathsf{W}(\cdot)$ denotes the Lambert W function, known as the inverse function of $f(x) = x\exp(x)$ \cite{LambertW}.
\item[2)] The asymptotically optimal network latency is given as
\begin{equation}\label{eq45}
\!\!\!{\mathcal{D}^ \star } = \left\{ {\begin{array}{*{20}{l}}
{\Theta \left( {\dfrac{\mathcal{T}_{\rm{th}}U}{{W\log \rho }}} \right),}&{\text{if Regime i)}},\\
{\Theta \left( {\dfrac{\mathcal{T}_{\rm{th}}U}{{W\sqrt M }}} \right),}&{\text{if Regime ii)}},\\
{\Theta \left( {\dfrac{\mathcal{T}_{\rm{th}}U}{{W{M^2}{\rho ^2}}}} \right),}&{\text{if Regimes iii) \& iv)}},
\end{array}} \right.
\end{equation}
\end{itemize}
\end{theorem}
\begin{IEEEproof}
See Appendix D.
\end{IEEEproof}


From Theorem 5, some implications can be discussed as follows:

\begin{itemize}
\item
\emph{\textbf{Regime i)}}: Here, the transmit energy constraint is sufficiently high and the number of BS antennas is not sufficiently high. So, this regime can be interpreted as the scenario that each BS equipped with a not-so-large number of antennas serves users with sufficient energy in a small-sized cell. In this case, the asymptotically optimal policy is to configure the half of each sub-frame as the training phase and to serve $N/2$ users simultaneously, which implies that orthogonal pilots are optimal. Note that this result is consistent to those in previous literature, such as {Theorem 2} in \cite{MurugesanMAC} and Sec. V in \cite{Hassibi}. Also, equal-energy allocation over all symbols, i.e., $$p^{\rm{tr},\star}_j =p^{\rm{dt},\star}_j =\dfrac{E_j}{N},$$ 
is nearly optimal, but the network latency cannot be reduced as the number of BS antennas increases because the growth rate of the BS antennas is too slow. In case $W=\Theta(U)$, the target throughput can be increased as $\Theta(\log\rho)$ while keeping the latency requirement, which is also consistent to the classical point of view.

\item
\emph{\textbf{Regime ii)}}: The transmit energy constraint and the number of BS antennas are both  sufficiently high. So, this regime can be interpreted as the scenario that each BS equipped with a very large number of antennas serves users with sufficient energy in a small-sized cell. In this case, it is asymptotically optimal to configure one-third of a sub-frame as the training phase and to serve $\approx0.3\sqrt{NM}$ users simultaneously for each sub-frame, which implies that non-orthogonal pilots become optimal. Also, the optimal energy allocation is given as 
\begin{align*}
p_j^{{\rm{tr}},\star} = \frac{{3\left( {2 - \sqrt 2 } \right){E_j}}}{{\sqrt 2 N}},~p_j^{{\rm{dt}},\star} = \frac{{3\left( {2 - \sqrt 2 } \right){E_j}}}{{2N}},
\end{align*}
which implies that although $1.5$dB higher energy per symbol is dedicated to each training symbol, but $1.5$dB higher energy is allocated to the data transmission phase.  
In this regime, the network latency can be arbitrarily reduced by increasing the number of BS antennas. However, allowing more energy is not beneficial. In case $W=\Theta(U)$, the target throughput can be increased as $\Theta(M^{1/2})$ while keeping the latency requirement or the latency is reduced as $\Theta(M^{-1/2})$ while keeping the target throughput.

\item
\emph{\textbf{Regime iii)}}: Here, the transmit energy constraint is quite tight and the number of BS antennas is not so high. This regime can be interpreted as the scenario that each BS equipped with a not-so-large number of antennas serves users with limited energy in a large-sized cell. In this case, it turns out that optimal pilots are non-orthogonal and the optimal energy allocation is
\begin{align*}
p_j^{{\rm{tr}},\star} &= \dfrac{{E_j}}{2L^\star},~
p_j^{{\rm{dt}},\star} = \dfrac{{E_j}}{{2\left( {N - L^\star} \right)}},
\end{align*}
which implies that the energy allocation is identical to that in Regime i), i.e., equal-energy allocation over all symbols becomes optimal if $L^\star =N/2$ is selected as in Region i). In case $W=\Theta(U)$, in order to meet the latency requirement, the target throughput needs to be scaled as $\Theta(\rho^2)$. 

\item
\emph{\textbf{Regime iv)}}: The transmit energy constraint is quite tight, but the number of BS antennas is sufficiently high. This regime can be interpreted as the scenario that each BS equipped with a very large number of antennas serves users with limited transmit energy in a large-sized cell. In this case, optimal pilots are non-orthogonal with the same asymptotically optimal scheduling policy to that for Regime iii), but with different optimal energy allocation is given as 
\begin{align*}
&p_j^{{\rm{tr}},\star} = \frac{{{E_j}}}{{\sqrt {L^\star\left( {N - L^\star} \right)}  + L^\star}},\\
&p_j^{{\rm{dt}},\star} = {\frac{{{E_j}}}{{\sqrt {L^\star\left( {N - L^\star} \right)}  + N - L^\star}}}.
\end{align*}
However, if $L^\star=N/2$ is selected as in Regime i), the equal-energy allocation becomes optimal. In case $W=\Theta(U)$, in order to meet the latency requirement, the target throughput needs to be scaled as $\Theta(M^2\rho^2)$.
\end{itemize}

\begin{remark}
It is worth noting that our analysis may be regarded as the results of the capacity-approaching receiver, even if we deal with simple linear receivers only because as $M/K\to \infty$, the lower-bound of the achievable rate of the ZF or MRC receiver converges to the exact achievable rate, i.e., $R_j\to\widetilde{R}_j$ and the achievable rate also converges to the capacity. The only required condition is $M/K \to \infty$, which is valid in Regimes i) and ii). However, this condition is not satisfied in Regimes iii) and iv), which implies that linear receivers become strictly sub-optimal if the network is operated in a limited energy regime.
\end{remark}

From the above, it is shown that orthogonal pilots become optimal only in Regime i), i.e., only  in a classical cellular system scenario, but non-orthogonal pilots become optimal in Regimes ii)-iv), i.e., in new scenarios for future cellular systems. To quantify the advantages of using non-orthogonal pilots, the asymptotically optimal network latency using orthogonal pilots only is given as follows.

\begin{corollary} Suppose that $\mathrm{Var}(E_j\beta_j)\ll\rho^2$ but the network does not allow non-orthogonal pilots (i.e., $K^\star \le L^\star$). As $U\to\infty$ and $M\to\infty$ with $U/M\to\infty$, the asymptotically optimal network latency is given as 
\begin{align}\label{eq46}
{\mathcal{D}^ \star } = \left\{ {\begin{array}{*{20}{l}}
{\Theta \left( {\dfrac{\mathcal{T}_{\rm{th}}U}{{W\log \rho }}} \right),}&{\text{if }\rho\gg 1 ~\text{and}~ M\ll \rho,}\\
{\Theta \left( {\dfrac{\mathcal{T}_{\rm{th}}U}{{W\log M}}} \right),}&{\text{if }\rho\gg 1~\text{and}~M\gg \rho,}\\
{\Theta \left( {\dfrac{\mathcal{T}_{\rm{th}}U}{{WM{\rho ^2}}}} \right),}&{\text{if }\rho \ll 1.}
\end{array}} \right.
\end{align}
\end{corollary}
\begin{IEEEproof}
Since non-orthogonal pilots are not allowed, the asymptotically optimal network latency is obtained by using $L^\star = 1$ and $K^\star = 1$. 
\end{IEEEproof}

As the number of BS antennas increases, the gain obtained by allowing non-orthogonal pilots becomes quite dramatic in Regimes ii)-iv). In Regimes iii) and iv), $M$ times lower network latency can be achieved. In Regime ii), the use of non-orthogonal pilots makes the network latency decrease sub-linearly rather than logarithmically with $M$. Thus, an important design guideline can be derived: \emph{for a latency-sensitive application, it would be better to serve more users by employing non-orthogonal pilots, which is quite dramatic in the case of being operated in a high energy regime with a high target throughput, and is still quite meaningful even in the case of being operated in a low energy regime}.

\begin{remark}
In order to prevent the misunderstanding on the above results, we would like to emphasize that each of users is assumed to have an independent transmit energy source for transmission in this paper. So, as the number of scheduled users increases, more energy is consumed at a given sub-frame. Thus, in a low energy regime, i.e., Regime iii) or iv), the gain coming from scheduling more users is larger than the channel estimation quality degradation, which is clearly different compared to the case in \cite{MurugesanMAC}, where the total energy dedicated for a transmission is fixed regardless of the number of scheduled users. 
\end{remark}

\section{Concluding Remarks}
In this paper, the latency-optimal static uplink scheduling policy is provided and its network latency is analyzed in an uplink training-based LSAS employing simple ZF or MRC receiver. The optimal uplink scheduling problem considered in this paper is to minimize the network latency when each user is constrained with a target throughput and energy limit and the corresponding optimal solution provides the optimal scheduling groups with their own scheduling portions, the optimal energy allocation between the training and data phases for each user, and the optimal frame configuration for the training based LSAS.
The optimal energy allocation is derived in a simple close-form for a given scheduling group and the optimal scheduling groups are found to be comprised of users with similar received signal quality. Then, a low-complexity uplink scheduling algorithm providing the exact optimal solution is proposed with polynomial-time complexity of $O(N(MU)^{3.5})$. Via numerical examples, it is shown that the proposed uplink scheduling algorithm can provide an optimal policy which can provide several times lower network latency at given throughput and energy constraints over the conventional non-optimized scheduling algorithms in realistic environments, which shows that the proposed work can be a key enabler for the oncoming 5G communication networks. 

In addition, the proposed uplink scheduling policy and the corresponding optimal network latency are analyzed asymptotically to provide better insights on the system behavior. As the number of users increases, it is shown that the network latency, normalized by the number of total users, converges to a deterministic quantity and asymptotically optimal 
frame configuration and scheduling policy can be obtained, which gives a way to construct much simpler asymptotically optimal uplink scheduling policy with complexity of $O(U\log U)$. Further, four operating regimes are classified according to the growth or decay rate of the average received signal quality and the number of BS antennas. It turns out that orthogonal pilot sequences, widely used in current systems, are optimal only when the average received signal quality is sufficiently good and the number of BS antennas is not-so-large, i.e., only in a conventional scenario. In other regimes representing new service and system scenarios, it turns out that non-orthogonal pilot sequences become optimal. By using non-orthogonal pilots, the network latency can be reduced by a factor of $\Theta(M)$ when the received signal quality is quite poor ($\rho\ll 1$) or by a factor of $\Theta(\sqrt{M}/\log M)$ when the received signal quality is sufficiently good ($\rho\gg 1$) and the number of BS antennas is sufficiently large ($M\gg \log_2\rho$). Thus, this work proves that, in order to minimize the network latency, it is better to schedule users more than the amount of available training resource by employing non-orthogonal pilots for a training-based LSAS, which would be a critical guideline for designing 5G cellular communication systems supporting massive MTC (or IoT) with low energy or latency-sensitive ultra-reliable Tactile Internet services.

\appendices
\section{Proof of Theorem 1}
For brevity, we consider the ZF receiver only and drop the indices $K$ and $L$. 
By taking the derivative of (\ref{eq29}), we obtain
\begin{equation}\label{eq47}
\begin{split}
 \frac{d}{{dx}}{\overline \Omega _q}(x;K,L)
&=\frac{{\left( {\left( {{a}{e} - {b}{d}} \right){x^2} - 2{b}{c}x + {a}{c}} \right){{\log }_2}e}}{{\left( {{c} + {d}x - {e}{x^2}} \right)\left( {{c} + \left( {{a} + {d}} \right)x - \left( {{b} + {e}} \right){x^2}} \right)}}.
\end{split}
\end{equation}
Since (\ref{eq29}) is non-negative and continuous on $[0,a/b]$ and $\overline{\Omega}_q(0;K,L)=\overline{\Omega}_q(a/b;K,L)=0$, $u^\star(\mathcal{O}_q)$ is obtained by finding a real root of the quadratic function $\frac{d}{dx}\overline{\Omega}_q(x;K,L)=0$ as long as (\ref{eq30}) is feasible, i.e., $0\le u^\star(\mathcal{O}_q)\le a/b$. First, show that $u^\star(\mathcal{O}_q)$ is real. To do this, it is sufficient that 
\begin{equation}\label{eq48}
\frac{a}{b}\frac{{{b}{d} - {a}{e}}}{{{b}{c}}} + 1 \ge 0.
\end{equation}
After inserting values in Table II, (\ref{eq48}) becomes equivalent to 
\[\left( {N - L} \right)\left( {\left( {{{\left( {K - L} \right)}^ + } + L} \right){E_K}{\beta _K} + L} \right) \ge 0,\]
which is always true since $0<L<N$. Now, show that $u^\star(\mathcal{O}_q)\ge 0$. If $bd-ae=0$, it is trivial so we omit. Suppose that $bd-ae<0$. Then, the condition $u^\star(\mathcal{O}_q)\ge 0$ can be written as
\[{u^\star}(\mathcal{O}_q) \ge 0 \Leftrightarrow \sqrt {1 + \frac{a}{b}\frac{{bd - ae}}{{bc}}}  \le 1,\]
which is also always true since $bd-ae<0$. Similarly $u^\star(\mathcal{O}_q)\ge 0$ when $bd-ae>0$.  Finally, show that $u^\star(\mathcal{O}_q)\le a/b$. The case ${b}{d} - {a}{e} = 0$ is again trivial so that we omit it. Assume that ${b}{d} - {a}{e} > 0$. Then, we have 
\[\sqrt {1 + \frac{{{a}}}{{{b}}}\frac{{{b}{d} - {a}{e}}}{{{b}{c}}}} \le 1+\frac{{{a}}}{{{b}}}\frac{{{b}{d} - {a}{e}}}{{{b}{c}}} ,\]
which holds since $\sqrt{1+x}\le 1+x$ for any $x\ge0$ and inserting it in (\ref{eq31}) shows $u^\star(\mathcal{O}_q)\le a/b$. Similarly for ${b}{d} - {a}{e} < 0$, we have 
\[\sqrt {1 + \frac{{{a}}}{{{b}}}\frac{{{b}{d} - {a}{e}}}{{{b}{c}}}} \ge 1 + \frac{{{a}}}{{{b}}}\frac{{{b}{d} - {a}{e}}}{{{b}{c}}},\] which completes the proof. 

\section{Proof of Theorem 2}
Without loss of generality, we assume $E_1\beta_1>E_2\beta_2>\dots>E_U\beta_U$ and $U\in \mathcal{O}^\star_{Q}$ and rewrite the objective function as 
\begin{align*}
&\sum\limits_{q = 1}^{Q} {{{(\Omega _q^ \star ({{\cal O}_q},L))}^{ - 1}}} = \sum\limits_{q = 1}^{Q-1} {{{(\Omega _q^ \star ({{\cal O}_q},L))}^{ - 1}}} + {(\Omega _{Q}^ \star ({{\cal O}_{Q}},L))^{ - 1}}.
\end{align*}
Note that $(\Omega_q^\star (\mathcal{O}_q,L))^{-1}$ is a monotonically decreasing function of $ E_{k_q}\beta_{k_q}$ for $k_q = \min_{j\in\mathcal{O}_q}E_j\beta_j$ and is independent to $E_{j}\beta_{j},~ \forall j\in \mathcal{O}_q\backslash\{k_q\}$. Since $U\in\mathcal{O}_Q^\star$, in order to minimize $\sum\nolimits_{q = 1}^{Q-1} {{{(\Omega _q^ \star ({{\cal O}_q},L))}^{ - 1}}}$, $\mathcal{O}^\star_{Q} $ should be 
$$\mathcal{O}^\star_{Q} = \{{U-|\mathcal{O}_{Q}|+1},{U-|\mathcal{O}_{Q}|}+2,\dots,U\}.$$
Similarly, $\mathcal{O}^\star_{q}$ is successively determined once $\mathcal{O}_r^\star$, $r= q+1,\dots,Q$,  are determined, which concludes that the two properties in Theorem 2 hold for all $\mathcal{O}_q^\star, \forall q$.

\section{Proof of Theorem 4}
\subsection{Preliminary}
Before proving Theorem 4, some preliminary results about a quantile function are introduced.

 Suppose that $X_1,\dots,X_n$ are i.i.d. real-valued random variables with CDF $F$ and the order statistics of $X_1,\dots,X_n$ are denoted by $X_{(1)},\dots,X_{(n)}$. For $0<p<1$, the $p$th quantile of $F$ is defined as $F^{-1}(p)= \xi_p = \inf\{x|F(x)\ge p\}$. Correspondingly, the sample quantile is defined as the $p$th quantile of the empirical CDF $F_n$ with $n$ samples, $F_n^{-1}(p) =\widehat{\xi}_p = \inf\{x|F_n(x)\ge p\}$, which can also be  expressed as $X_{(\lceil np\rceil)}$.
\begin{lemma}
Let $X_1,\dots, X_n$ be i.i.d. random variables from a CDF $F$ satisfying $p < F(\xi_p+\epsilon)$ for any $\epsilon> 0$. Then, for every $\epsilon > 0$ and $n = 1, 2,\dots$,
\begin{equation}\label{eq49}
\Pr(|\widehat{\xi}_p-\xi_p|>\epsilon)\le 2Ce^{-2n\delta_\epsilon^2},
\end{equation}
where $\delta_{\epsilon} = \min\{F(\xi_p+\epsilon)-p,p-F(\xi_p-\epsilon)\}$ and $C$ is a positive constant.
\end{lemma}
\begin{IEEEproof}The proof is directly obtained by applying the Dvoretzky-Kiefer-Wolfowitz inequality \cite{VaartAsymptotic}.
\end{IEEEproof}

Now, show the almost-sure convergence of $\widehat{\xi}_p$ as $n\to\infty$. 
\begin{lemma}
Let $X_1,\dots, X_n$ be i.i.d. random variables from a CDF $F$. Then, $\widehat{\xi}_p\xrightarrow{a.s.}\xi_p$.
\end{lemma}
\begin{IEEEproof}
For any given $\epsilon>0$ and sufficiently large $N$, we have 
\begin{equation*}
\begin{split}
\Pr \left( {\left| {{{\widehat \xi }_p} - {\xi _p}} \right| > \epsilon,\text{some } n >N} \right) &= 
\Pr \left( {\bigcup\nolimits_{n = N+1}^\infty  {\left\{ {\left| {{{\widehat \xi }_p} - {\xi _p}} \right| > \epsilon} \right\}} } \right) \\
&\le \sum\nolimits_{n = N+1}^\infty  {\Pr \left( {\left| {{{\widehat \xi }_p} - {\xi _p}} \right| > \epsilon} \right)} \\
&\le 2C\sum\nolimits_{n = N+1}^\infty  {{e^{ -2n\delta_\epsilon^2}}},
\end{split}
\end{equation*}
which can be made arbitrarily small by increasing $N$ because $\sum_{n=1}^\infty e^{-2n\delta_\epsilon^2}$ is convergent. Thus, $\mathop {\lim }\nolimits_{n \to \infty } \Pr \left( {\left| {{{\widehat \xi }_p} - {\xi _p}} \right| \le\epsilon} \right) \to 1$ as $n\to\infty$, which implies $\widehat{\xi}_p\xrightarrow{a.s.}\xi_p$.
\end{IEEEproof}

\begin{lemma}
For a Riemann-integrable function $g$, we have 
\begin{equation}\label{eq50}
\frac{1}{n}\sum\limits_{i = 1}^n {g\left( {{X_{(i)}}} \right)}  \xrightarrow{a.s.} \int_0^1 {g\left( {{F^{ - 1}}\left( t \right)} \right)dt}.
\end{equation}
\begin{IEEEproof}
From Lemma 7, as $n\to\infty$,
\begin{align*}
\frac{1}{n}\sum\limits_{i = 1}^n {g\left( {{X_{(i)}}} \right)}  &\xrightarrow{a.s.} \frac{1}{n}\sum\limits_{i = 1}^n {g\left( {{F^{ - 1}}\left( {\frac{i}{n}} \right)} \right)} \\
 &\to \int_0^1 {g\left( {{F^{ - 1}}\left( t \right)} \right)dt},
\end{align*}
where the last convergence comes from the definition of the Riemann integral. 
\end{IEEEproof}
\end{lemma}

\begin{lemma}
Let $X_{(i_1)},\dots,X_{(i_Q)}$ be the $Q$ samples of $U$ i.i.d random variables $X_{(1)},\dots,X_{(U)}$ with $K_q=i_{q+1}-i_{q}$. Define $K = {U}/{Q}$ and assume that $\max_{q=1,\dots,Q} K_q/U\to 0$ as $U\to\infty$. Then, 
\begin{equation}\label{eq51}
\mathop {\lim }\limits_{Q \to \infty } \left| {\frac{1}{Q}\sum\limits_{q = 1}^Q {g\left( {{X_{\left( {qK} \right)}}} \right)}  - \frac{1}{Q}\sum\limits_{i = 1}^Q {g\left( {{X_{\left( {{i_q}} \right)}}} \right)} } \right| \xrightarrow{a.s.} 0.
\end{equation}
\end{lemma}
\begin{IEEEproof}
Since each $K_q/U\to0$ and $\sum_{q=1}^Q K_q = U$, $Q\to\infty$ as $U\to\infty$.
From Lemma 8, we have 
\begin{align*}
\frac{1}{Q}\sum\limits_{q = 1}^Q {g\left( {{X_{\left( {qK} \right)}}} \right)} &\xrightarrow{a.s.} \frac{1}{Q}\sum\limits_{i = 1}^Q {g\left( {{F^{ - 1}}\left( {\frac{i}{Q}} \right)} \right)} \\
& \to \int_0^1 {g\left( {{F^{ - 1}}\left( t \right)} \right)dt}.
\end{align*}
Since $F^{-1}$ is continuous and each $K_q/U\to0$, small variation in the length of each interval ($K_q/U$ vs. $K/U = \frac{1}{QU} \sum_{q=1}^Q K_q$) does not affect the convergence of the Riemann integral so that 
\begin{align*}
\frac{1}{Q}\sum\limits_{q = 1}^Q {g\left( {{X_{\left( {{i_q}} \right)}}} \right)} &\xrightarrow{a.s.} \frac{1}{Q}\sum\limits_{q = 1}^Q {g\left( {{F^{ - 1}}\left( {\frac{{{i_q}}}{U}} \right)} \right)} \\
 &\to \int_0^1 {g\left( {{F^{ - 1}}\left( t \right)} \right)dt},
\end{align*}
which concludes the proof.
\end{IEEEproof}

\subsection{Proof of Theorem 4}

 In this proof, we omit the index $L$ by assuming $L^\star$ is used in the symbols $a_{K,L}$, $b_{K,L}$, $c_{K,L}$, $d_{K,L}$, and $e_{K,L}$ for simplicity. 
Assume that $E_1\beta_1\ge E_2\beta_2 \ge \dots \ge E_U\beta_U$ and let $\{K_i\}$ be the sequence of 
positive finite integers such that $K_1+K_2+\dots+K_Q = U$ and $\mathcal{O}^\star_q=\{i^q_{1},\dots,i^q_{K_q}\}$, $\forall q$, where $i^q_{k} = K_1+\dots+K_{q-1} + k$ for $k=1,\dots,K_q$.

We first prove that the optimal scheduling groups can be selected among equi-sized ones as $U\to\infty$. 
From (\ref{eq23}) and (\ref{eq24}), we have 
\begin{equation}\label{eq52}
{\mathcal{D}^ \star } = \frac{{\eta {\mathcal{T}_{{\rm{th}}}}}}{W}\frac{Q}{{1 - \frac{L}{N}}}\Lambda,
\end{equation}
where $Q$ is the number of scheduling groups ($Q\to\infty$ as $U\to\infty$) and 
\begin{equation}\label{eq53}
\Lambda={{\frac{1}{Q}\sum\limits_{q=1}^Q {{{\left( {\Omega _q^\star}(\mathcal{O}_q^\star,L^\star)\right)}^{ - 1}}} }}.
\end{equation} 
Note that in (\ref{eq29}),  $\Omega_q^\star(\mathcal{O}_q;L^\star)$ is given as 
\begin{equation}\label{eq54}
\begin{split}
&\Omega _q^ \star ({{\cal O}_q};L^\star) ={\log _2}\left( {1 + \frac{{\left({a_{{i^q_{{K_q}}}}} - {b_{{i^q_{{K_q}}}}}{u^ \star }({{\cal O}_q},L^\star)\right){u^ \star }({{\cal O}_q},L^\star)}}{{{c_{{i^q_{{K_q}}}}} + {d_{{i^q_{{K_q}}}}}{u^ \star }(\mathcal{O}_q,L^\star) - {e_{{i^q_{{K_q}}}}}{{\left( {{u^ \star }({{\cal O}_q},L^\star)} \right)}^2}}}} \right).
\end{split}
\end{equation}
Since $u^\star(\mathcal{O}_q;L^\star)$ depends only on $E_{{{i^q_{{K_q}}}}}\beta_{{{i^q_{{K_q}}}}}$, denote
\[ (\Omega_q^\star(\mathcal{O}_q^\star,L^\star))^{-1}=\upsilon \left( {{{E_{{i^q_{{K_q}}}}}{\beta _{{i^q_{{K_q}}}}}}} \right)\]
by allowing some notational abuse. From Lemma 9 as $U\to\infty$, 
\[\mathop {\lim }\limits_{Q \to \infty } \left| {\frac{1}{Q}\sum\limits_{q = 1}^Q {\upsilon \left( {{E_{i_{{K_q}}^q}}{\beta _{i_{{K_q}}^q}}} \right)}  - \frac{1}{Q}\sum\limits_{i = 1}^Q {\upsilon\left( {{E_{q{K^*}}}{\beta _{q{K^*}}}} \right)} } \right|\xrightarrow{a.s.}0.\]
which implies that the equi-sized scheduling group with size of $K^\star$ can achieve the optimal network latency asymptotically. Also from Lemma 8, the asymptotically optimal network latency can be expressed as
\begin{equation}\label{eq55}
\begin{split}
\Lambda &\xrightarrow{a.s.} \frac{1}{Q}\sum\limits_{q = 1}^Q {v\left( {{E_{q{K^\star}}}{\beta _{q{K^\star}}}} \right)} \\
 &\to \int_0^1 {v\left( {{F^{ - 1}}(t)} \right)dt}.
\end{split}
\end{equation}
 First consider the case of $K^\star\le L^\star$ and $N= L^\star+K^\star$ so that $b_{K^\star}d_{K^\star}-a_{K^\star}e_{K^\star}$ = 0. Then, $u^\star (\mathcal{O}^\star_q)$ is given as
\begin{equation}\label{eq56}
u^\star(\mathcal{O}^\star_q,L^\star) = \dfrac{a_{{{{qK^\star}}}}}{2b_{{{{qK^\star}}}}} = \dfrac{E_{{{{qK^\star}}}}\beta_{{{{qK^\star}}}}}{2L^\star}.
\end{equation}
Inserting (\ref{eq56}) into (\ref{eq54}), we obtain
\begin{equation}\label{eq57}
\begin{split}
&\log _2\left( {1 + \frac{{a_{qK^\star}^2}}{{4{b_{qK^\star}}{c_{qK^\star}}}}} \right) = {\log _2}\left( {1 + \frac{{\left( {M - K^\star} \right){{\left( {{E_{qK^\star}}{\beta _{qK^\star}}} \right)}^2}}}{{4K^\star\left( {1 + {E_{qK^\star}}{\beta _{qK^\star}}} \right)}}} \right).
\end{split}
\end{equation}
By using (\ref{eq55}), we obtain
\begin{equation}\label{eq58}
\begin{split}
\Lambda &\xrightarrow{a.s.} \int_0^1 {\log _2^{ - 1}\left( {{\rm{1 + }}\frac{{\left( {M - K^\star} \right){{\left( {{F^{ - 1}}(t)} \right)}^2}}}{{4K^\star\left( {1 + {F^{ - 1}}(t)} \right)}}} \right)dt}  \\
&= \int_0^\infty  {\log _2^{ - 1}\left( {{\rm{1 + }}\frac{{\left( {M - K^\star} \right){x^2}}}{{4K^\star\left( {1 + x} \right)}}} \right)f(x)dx},
\end{split}
\end{equation}
which concludes the proof for the case. The other cases can be shown in similar ways.

\section{Proof of Theorem 5}
\subsection{Preliminary}
\begin{lemma}
For $x\gg 1$, 
\begin{align}\label{eq59}
h(x;L,K) \cong  \left\{ {\begin{array}{*{20}{l}}
{\frac{{\left( {M - K} \right)L}}{{K(K - L)}},}&{{\text{if }}K > L,}\\
{\frac{{M - K}}{{N - L + K + 2\sqrt {K\left( {N - L} \right)} }}x,}&{{\text{if }}K \le L,}
\end{array}} \right.
\end{align}
and for $x\ll1$, 
\begin{align}\label{eq60}
h(x;L,K) \cong \frac{{M - K}}{{4\left( {N - L} \right)}}{x^2}.
\end{align}
\end{lemma}
\begin{IEEEproof}
For $x\gg 1$, we obtain
\begin{align*}
&h(x;L,K) \cong \left\{ {\begin{array}{*{20}{l}}
{\frac{{\left( {M - K} \right)L}}{{K(K - L)}},}&{{\text{if }}K > L,}\\
{\frac{{\left( {M - K} \right)\left( {N - L + K - 2\sqrt {K\left( {N - L} \right)} } \right)}}{{{{\left( {N - L - K} \right)}^2}}}x,}&{{\text{if }}K \le L,~N \ne L + K,}\\
{\frac{{M - K}}{{4K}}x,}&{{\text{if }}K \le L,~N = L + K.}
\end{array}} \right.
\end{align*}
Using the equality
\begin{equation}
\begin{split}\label{eq61}
{{{\left( {N - L - K} \right)}^2}}&{ = {N^2} - 2N\left( {L - K} \right) + {{\left( {L - K} \right)}^2} - 4K\left( {N - L} \right)}\\
&{ = \left( {N - L + K - 2\sqrt {K\left( {N - L} \right)} } \right)}
{ \left( {N - L + K + 2\sqrt {K\left( {N - L} \right)} } \right)}
\end{split}
\end{equation}
simplifies the case $K\le L$, $N\ne L+K$, and  inserting $N={L+K}$ into (\ref{eq61}) results in 
\[N - L + K + 2\sqrt {K\left( {N - L} \right)} = 4K,\]
by which we arrive at (\ref{eq59}). On the other hand, by using Taylor expansion, $1-\sqrt{(ax+1)(bx+1)} = -\frac{1}{2}(a+b)x+\frac{1}{8}(a-b)^2x^2+O(x^3)$ for $x\ll1$. Then, for $x\ll 1$, we obtain
\begin{equation*}
\begin{split}
&h(x;L,K) \cong \left\{ {\begin{array}{*{20}{l}}
{\frac{{\left( {M - K} \right)}}{{4\left( {N - L} \right)}}{x^2},}&{{\text{if }}K > L,}\\
{\frac{{\left( {M - K} \right){{\left( {K\sqrt {\frac{L}{{N - L}}} - L\sqrt {\frac{{N - L}}{L}} } \right)}^2}}}{{4L{{\left( {N - L - K} \right)}^2}}}{x^2},}&{{\text{if }}K \le L,~N \ne L + K,}\\
{\frac{{\left( {M - K} \right)}}{{4K}}{x^2},}&{{\text{if }}K \le L,~N = L + K.}
\end{array}} \right.
\end{split}
\end{equation*}
To simplify the case $K\le L, N\ne {L+K}$, we use the identity
\begin{equation}\label{eq62}
{\left( {K\sqrt {\frac{L}{{N - L}}} - L\sqrt {\frac{{N - L}}{L}} } \right)^2} =\frac{ L{\left( {N - L - K} \right)^2}}{{N - L}},
\end{equation}
by which we arrive at (\ref{eq60}).
\end{IEEEproof}

\begin{lemma}
Let $\{X_n\}$ be a sequence of positive random variables. Suppose that $\mathrm{Var}(X_n) =o((\mathbb{E}[X_n])^2)$. Then, for any continuous function $g$, $g(X_n) = \Theta(\mathbb{E}[X_n])$ almost-surely.
\end{lemma}
\begin{IEEEproof}
By using the Chebyshev's inequality, we obtain, as $n\to\infty$
\begin{equation*}
{\rm{Pr}}\left( {\left| {{{\rm{X}}_n} - \mathbb{E}[{X_n}]} \right| > c\mathbb{E}[{X_n}]} \right) \le \frac{{{\rm{Var}}\left( {{X_n}} \right)}}{{{c^2}{{\left( {\mathbb{E}[{X_n}]} \right)}^2}}} \to 0,
\end{equation*}
where $c$ is a finite and positive constant, which implies that the realizations of $X_n$ is included in the set $\mathcal{X}_n=\{X_n|(1-c)\mathbb{E}[X_n]\le X_n\le(1+c)\mathbb{E}[X_n]\}$ almost-surely. Thus, the minimum and maximum in $\mathcal{X}_n$ is scaled as $\mathbb{E}[X_n]$ almost-surely so that its transformation, i.e., $g(X_n)$ is also scaled as $g(\mathbb{E}[X_n])$ almost-surely by continuous mapping theorem \cite{VaartAsymptotic}, which completes the proof.
\end{IEEEproof}

\subsection{Proof of Theorem 5}

{\emph{Regime i)}}: Consider the case $K\le L$ and let $a = {\frac{{M - K}}{{N - L + K + 2\sqrt {K\left( {N - L} \right)} }}}$. We Obtain
\begin{align}\nonumber
\mathbb{E}\left[ {\frac{1}{{{{\log }_2}(1 + aX)}}} \right] & = \Theta \left( {\frac{1}{{\log \left( {1 + a{\rho}} \right)}}} \right)\\
& = \Theta \left( {\frac{1}{{\log {\rho}}}} \right),\label{eq63}
\end{align}
where the first equality comes from Lemma 11 and the last equality comes from $\rho \gg a$. So, $\mathbb{E}\left[ {\log _2^{ - 1}\left(1+ aX \right)} \right]$ becomes independent to $K$ and $L$ asymptotically. By using Theorem 5 and (\ref{eq63}), we obtain
\begin{align}\nonumber
( {L^\star},{K^\star} )&= \mathop {\arg \min }\limits_{K \le L } \frac{\mathbb{E}\left[ {\log _2^{ - 1}\left(1+ aX \right)} \right]}{{K\left( {1 - \frac{L}{N}} \right)}} 
\\&=\mathop {\arg \max }\limits_{K \le L} K\left( {1 - \frac{L}{N}} \right) \nonumber
\\&= \left(\frac{N}{2},\frac{N}{2}\right),\label{eq64}
\end{align}
and by inserting (\ref{eq64}) into (\ref{eq43}), we obtain
\begin{align}\nonumber
{\overline {\mathcal{D}} ^\star} &= \frac{\eta\mathcal{T}_{\rm{th}}}{W}\frac{4}{N}\mathbb{E}\left[ {\log _2^{ - 1}\left(1+ {\frac{{M - K}}{{2N}}X} \right)} \right]\\
&= \Theta \left( \frac{\mathcal{T}_{\rm{th}}}{W{\log \rho }} \right),\label{eq65}
\end{align}
where the last equality comes from Lemma 11 and $\rho\gg M$.

Now, consider the case $K> L$. By using Lemma 10, we can rewrite (\ref{eq41}) as 
\begin{align*}
\left( {{L^\star},{K^\star}} \right) &= \mathop {\arg \max }\limits_{L<K} {\nu(L,K)},
\end{align*}
where 
\begin{equation}\label{eq66}
\nu(L,K) = K\left( {1 - \frac{L}{N}} \right){\log _2}\left(1+\frac{{\left( {M - K} \right)L}}{{K(K - L)}}\right).
\end{equation}
If $M$ is finite, $K$ is also finite so that $\nu(L^\star,K^\star) = \Theta(1)$ is maximized at $L^\star = \Theta(1)$ and $K^\star = \Theta(1)$. Inserting it into (\ref{eq43}), we obtain 
\begin{equation}\label{eq67}
{\overline {\mathcal{D}} ^\star} = \frac{\eta\mathcal{T}_{\rm{th}}}{W}\frac{1}{\nu(L^\star, K^\star)} = \Theta\left(\frac{\mathcal{T}_{\rm{th}}}{W}\right).
\end{equation}
Comparing (\ref{eq67}) with (\ref{eq65}) indicates that the asymptotically lower network latency can be achieved when $L\ge K$. Now, consider the case when $M\gg 1$. For a fixed $L$ and $M\gg 1$, $\nu(L,K)$ can be approximated as 
\begin{equation}\label{eq68}
\nu(L,K) \cong K\left( {1 - \frac{L}{N}} \right){\log _2}\left(1+\frac{{ {M} L}}{{K(K - L)}}\right),
\end{equation}
which is maximized at $K^\star = \chi\sqrt{M} + o(\sqrt{M})$ for some positive $\chi$. Ignoring the non-dominant term $o(\sqrt{M})$, inserting $K = \chi\sqrt{M}$, and replacing $L$ with $\lambda\in[1,K]$ in (\ref{eq68}) yields
\begin{equation}\label{eq69}
\overline{\nu} (\lambda, \chi) = \chi \sqrt M \left( {1 - \frac{\lambda}{N}} \right)\log _2\left(1+\frac{\lambda}{{{\chi ^2}}}\right),
\end{equation}
by which we obtain optimal $(\lambda^\star, \chi^\star)$ instead of using (\ref{eq68}). Since (\ref{eq69}) is unimodal, it is sufficient to find the point $(\lambda^\star,\chi^\star)$ satisfying ${{\frac{\partial }{\partial \lambda}}\left. {\overline{\nu} (\lambda,\chi)} \right|_{\lambda = {\lambda^\star}}} = 0$ and ${ \frac{\partial }{{\partial \chi }}\left.{\overline{\nu} (\lambda,\chi)} \right|_{\chi = {\chi ^\star}}} = 0$, which are respectively given as
\begin{align}\label{eq70}
&\log \left( {1 + \frac{\lambda^\star}{{{(\chi^\star) ^2}}}} \right) - \frac{{2\lambda^\star}}{{{(\chi^\star) ^2} + \lambda^\star}} = 0,\\
&\log \left( {1 + \frac{\lambda^\star}{{{(\chi^\star) ^2}}}} \right) - \frac{{N - \lambda^\star}}{{{(\chi^\star) ^2} + \lambda^\star}} = 0. \label{eq71}
\end{align}
Subtracting (\ref{eq70}) from (\ref{eq71}) results in $\lambda^\star = \frac{N}{3}$ and $\chi^\star $ should satisfy
\begin{equation*}
\log \left( {1 + \frac{1}{v}}\right) = \frac{2}{{v + 1}},
\end{equation*}
for  $v={{3{(\chi^\star) ^2}}}/N$. After some mathematical manipulations, we arrive at 
\[
-2e^{-2} = -2\frac{v}{v+1}e^{-2\frac{v}{v+1}},
\]
from which we have $\mathsf{W}(-2e^{-2}) = -2v/(v+1)$ and thus $v = 2/(\mathsf{W}(2e^{-2})+2)-1$ by introducing the Lambert W function. By inserting $(L^\star,K^\star) = (N/3,\chi^\star\sqrt{M}+o(\sqrt{M}))$ in (\ref{eq43}), we obtain
\begin{align}\nonumber
{\overline {\mathcal{D}} ^\star} &= \frac{\eta\mathcal{T}_{\rm{th}}}{W}\frac{1}{{\frac{2}{3}{\chi ^\star}\sqrt M {{\log }_2}\left( {\frac{1}{{{{\left( {{\chi ^\star}} \right)}^2}}}\frac{N}{3}} \right)}+o(\sqrt{M})}\\
 &= \Theta \left( \frac{\mathcal{T}_{\rm{th}}}{W\sqrt{M}}\right).\label{eq72}
\end{align}
Combining (\ref{eq65}) and (\ref{eq72}), we obtain $\overline{\mathcal{D}}^\star = \Theta(\frac{\mathcal{T}_{\rm{th}}}{W\log\rho})$.

{\emph{Regime ii)}}: Consider the case $M\ll \rho$. Then, we obtain both (\ref{eq65}) and (\ref{eq72}). Now, consider the case $M\gg \rho$. For $K>L$, we still obtain (\ref{eq72}).  However, for $K\le L$, $aX=\Theta(M\rho)$ so that $\mathbb{E}[\frac{1}{\log_2(1+aX)}] = \Theta(\frac{1}{\log M})$. Although (\ref{eq63}) still holds, (\ref{eq65}) becomes $\Theta(\frac{\mathcal{T}_{\rm{th}}}{W\log M})$, which, together with (\ref{eq72}), results in $\overline{\mathcal{D}}^\star = \Theta(\frac{\mathcal{T}_{\rm{th}}}{W\sqrt{M}})$.

{\emph{Regime iii) and iv)}}: From Lemma 10, (\ref{eq41}) can be rewritten as 
\begin{align}\nonumber
\left( {{L^\star},{K^\star}} \right) &= \mathop{\arg\min}\limits_{\scriptstyle 1\le L<N,\atop \scriptstyle 1\le K\le M} \frac{{\mathbb{E}\left[ {\log _2^{ - 1}\left( {1 + \frac{{\left( {M - K} \right)}}{{4\left( {N - L} \right)}}{X^2}} \right)} \right]}}{{K\left( {1 - \frac{L}{N}} \right)}}\\
 &= \mathop{\arg\min}\limits_{{\scriptstyle 1\le L<N,\atop \scriptstyle 1\le K\le M}} \frac{{4N{{\log }_2}e}}{{K\left( {M - K} \right)}}\mathbb{E}\left[ {{X^{ - 2}}} \right],\label{eq73}
\end{align}
where the second equality comes from that $\log_2(1+x) = x\log_2 e $ for small $x$. Finally, $K^\star = M/2$ and $L^\star$ can be selected arbitrarily among integers between $1$ and $N-1$. Inserting $(L^\star,K^\star) = (1, M/2)$, we obtain $\overline{\mathcal{D}}^\star = \Theta(\frac{\mathcal{T}_{\rm{th}}}{W(M\rho)^2})$, which competes the proof.

\end{document}